\documentclass[%
reprint,
 amsmath,amssymb,
 aps,
prb,
floatfix,
]{revtex4-2}
\newcommand\varpm{\mathbin{\vcenter{\hbox{%
  \oalign{\hfil$\scriptstyle+$\hfil\cr
          \noalign{\kern-.3ex}
          $\scriptscriptstyle({-})$\cr}%
      }}}}    
\usepackage{graphicx}
\usepackage{dcolumn}
\usepackage{bm}
\usepackage{physics}
\usepackage{xcolor}
\usepackage{gensymb}
\usepackage[caption=false]{subfig}
\newcommand{\sign}{\text{sgn}}

\usepackage{todonotes}

\begin{document}

\title{Two-dimensional altermagnets: Superconductivity in a minimal microscopic model}

\author{Bj\o rnulf Brekke, Arne Brataas and Asle Sudb\o\ }
\affiliation{Center for Quantum Spintronics, Department of Physics, Norwegian University of Science and Technology, NO-7491 Trondheim, Norway}

\date{\today}

\begin{abstract}
We propose a minimal toy model for a two-dimensional altermagnet. The model unravels altermagnetic properties at a microscopic level. We find spin-split electron- and non-degenerate magnon bands with a $d$-wave symmetry. We use the model to explore magnon-mediated superconductivity in altermagnets. The dominant superconducting state is spin-polarized with a $p$-wave symmetry. The state adopts its characteristics from the spin-split electron bands. Furthermore, we find that the superconducting critical temperature of altermagnets can be significantly enhanced by tuning the chemical potential.
\end{abstract}

\maketitle


\section{\label{sec:introduction}Introduction}
Altermagnets constitute a new subclass of magnetic materials \cite{SmejkalPRX2022}. They are defined as compensated collinear magnets with magnetic sublattices related by rotations instead of inversion or translation \cite{SmejkalSinovaPRX2022}. The symmetry of altermagnets allows for effects with a potential to rapidly progress the fields of spintronics \cite{HernandezPRL2021, SmejkalPRX22Hellenes, bose2022tilted, BaiPRL2022, KarubePRL2022, shao2021spin}, spin caloritronics \cite{zhou2023crystal} and superconductivity \cite{mazin2022notes, zhu2023topological, ouassou2023josephson, sun2023andreev, papaj2023andreev, zhang2023finite, ghorashi2023altermagnetic}. The most prominent feature of altermagnets is the large momentum-dependent spin split electron bands \cite{YuanPRB2020, fedchenko2023observation, AhnPRB2019, YuanPRM2021, reichlova2020macroscopic, noda2016momentum, hayami2019momentum, egorov2021colossal, BrekkePRB2022}. Nevertheless, the properties of altermagnets go beyond their characteristic alternating electron bands and include \textit{e.g.} chiral magnons \cite{vsmejkal2022chiral}. 

Spin splitting in altermagnets does not rely on a sizable spin-orbit coupling. This allows for a diverse range of potential materials. So far, RuO$_2$ has been the main focus of attention \cite{vsmejkal2020crystal, feng2022anomalous}, but there are many candidate materials \cite{reichlova2020macroscopic, mazin2021prediction, SmejkalSinovaPRX2022, MakotoPRB2021, YuanPRB2020, noda2016momentum, YuanPRM2021, lopez2016first, okugawa2018weakly, guo2023spin, ma2021multifunctional}. These systems are three-dimensional or quasi-two-dimensional. A monolayer altermagnet would be compelling in the rapidly developing field of van der Waals materials and other two-dimensional magnets \cite{burch2018magnetism, huang2017layer, geim2013van}. One route to altermagnetism is an anisotropic ordering of local orbitals \cite{naka2019spin}. However, the recent large experimental interest and progress has mainly been on materials where the altermagnetic properties rely on the interplay between magnetic and non-magnetic atoms. This typically calls for more involved unit cells than ferro- and antiferromagnets.
Consequently, much of the research on altermagnets carried out so far has been phenomenological \cite{SmejkalSinovaPRX2022, PengfeiPRX2022, bhowal2022magnetic, mcclarty2023landau}, through \textit{ab initio} calculations or experimental work. The microscopic models for studying altermagnets have typically been effective spin-dependent hopping models in momentum space that replicate electron bands with $d$-wave symmetry. Although useful, these models do not capture the magnetic ordering, and thus, the origin of the altermagnetic properties remains hidden from a microscopic point of view.

The interplay of magnetism and superconductivity has been a fertile playground for intriguing physics \cite{linder2015superconducting, BergeretRMP2005, robinson2010controlled, eschrig2015spin}.
The relation between the unconventional spin-splitting in altermagnets and unconventional superconductivity is an open question. This relation is fascinating because La$_2$CuO$_4$, which is a parent compound of high-temperature superconductors \cite{bednorz1986possible}, is an altermagnet \cite{ChakravartyPRB1989, SmejkalSinovaPRX2022}.
One intriguing direction is how the fluctuations in localized spins can mediate effective electron interactions that, in turn, give rise to superconductivity \cite{KargarianPRL2016, RohlingPRB2018, FjarbuPRB2019, MaelandPRL2023}.  

Here, we propose a two-dimensional minimal model to study altermagnets microscopically. The model sheds light on non-magnetic sites as a microscopic origin of the unconventional $d$-wave spin-split bands and non-degenerate magnon modes. We find that the interplay between electrons and magnons gives rise to a spin-polarized $p$-wave superconducting state, in which the critical temperature is tunable by the chemical potential. For specific values, we find a dramatic increase in the critical temperature analogous to the squeezing enhancement in conventional antiferromagnets \cite{Erlandsen2019PRB, Erlandsen2021PRB}, however, without the need for an uncompensated interface.

\begin{figure}[t!]
\includegraphics[width=0.75\columnwidth]{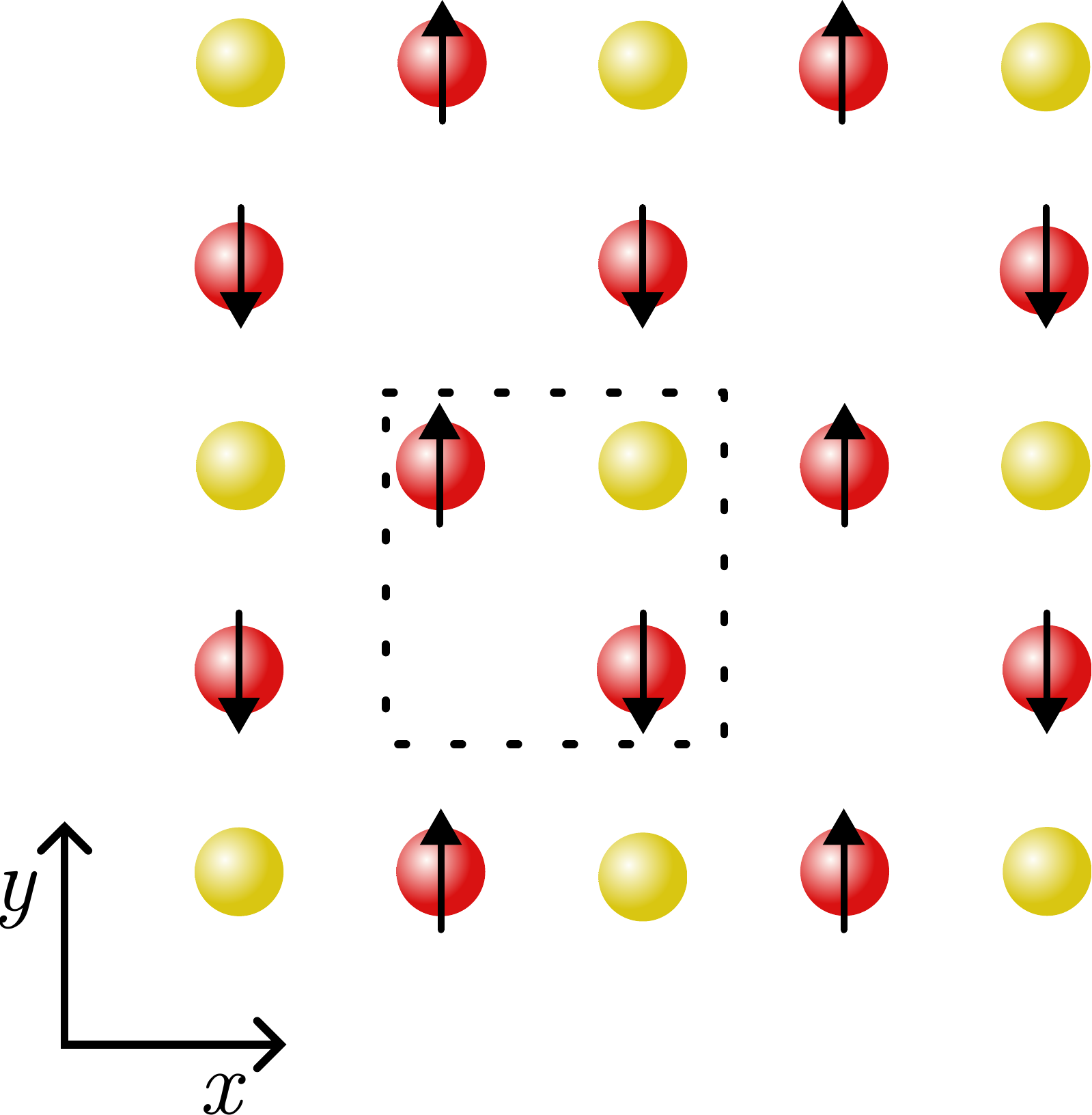}
\caption{\label{fig:lattice}A two-dimensional altermagnetic crystal lattice. The unit cell consists of three distinct sites for which rotational symmetries relate the two magnetic sites.}
\end{figure}

\section{\label{sec:Model}Model}
We consider the lattice in Fig. \ref{fig:lattice}. The non-magnetic lattice belongs to the plane group $p4mm$ (\#11).

The magnetic plane point group of the crystal has eight symmetries. These are
\begin{subequations}
\label{eq:whole}
\begin{align}
    (E\vert0), && (C_{2z}\vert0), && (\sigma_{x}\vert0), && (\sigma_{y}\vert0),
    \label{eq:sym1}
\end{align}
\begin{align}
    (C^+_{4z}\vert\mathcal{T}), && (C^-_{4z}\vert\mathcal{T}), && (\sigma_{xy}\vert\mathcal{T}), && (\sigma_{x\bar{y}}\vert\mathcal{T}),
    \label{eq:sym2}
\end{align}
\end{subequations}
where $\mathcal{T}$ is the time-reversal operator. The operations $C^-_{4z}$ and $C^+_{4z}$ are fourfold clockwise and counter-clockwise rotations about the $z$-axis. Furthermore, $\sigma_x$, $\sigma_y$, $\sigma_{xy}$, and $\sigma_{x\bar{y}}$ are mirror operations about the $x$-axis, $y$-axis and the diagonals.
The lattice is centrosymmetric because the two-fold rotation $C_{2z}$ acts as inversion symmetry in two dimensions. However, importantly, this inversion symmetry does not relate the two magnetic sublattices. Instead, they are related by the mirror- and rotational symmetries in Eq. \eqref{eq:sym2}. Thus, the magnetic crystal classifies as an altermagnet. In Appendix \ref{sec:orbitalOrdering}, we briefly consider an alternative altermagnetic model due to an anisotropic ordering of local orbitals.

\section{\label{sec:electrons}Electron Properties}
To study the electron properties, we employ a tight-binding Hamiltonian
\begin{align}
\begin{split}
    H_e = t\sum_{\left<i,j\right>,\sigma}  c_{i,\sigma}^{\dagger} c_{j,\sigma}  + t_2\sum_{\left<\left<i,j\right>\right>, \sigma}  c_{i,\sigma}^{\dagger} c_{j,\sigma} \\ 
  - J_{\mathrm{sd}}\sum_{i, \sigma, \sigma'} \bm{S}_i\cdot c_{i,\sigma}^{\dagger}\bm{\sigma}_{\sigma \sigma'} c_{i,\sigma'} -\mu \sum_{i, \sigma}  c_{i,\sigma}^{\dagger} c_{i,\sigma} \\ + \varepsilon_{\mathrm{nm}} \sum_{i\in \mathrm{nm}, \sigma} c_{i,\sigma}^{\dagger} c_{i,\sigma},
    \label{electronHamiltonian}
\end{split}
\end{align}
where $c_{i,\sigma}^{(\dagger)}$ is the (creation) annihilation operator of an electron at site $i$ with spin $\sigma$. The nearest-neighbor hopping strength is denoted by $t$. The next-nearest-neighbor hopping strength $t_2$ governs the hopping between the magnetic sites with opposite magnetization. The chemical potential $\mu$ determines the overall doping level, and  $\varepsilon_{\mathrm{nm}}$ is the non-magnetic site energy. We set the magnetic site energy to zero. The coupling $J_{\mathrm{sd}}$ quantifies the onsite exchange interaction between the spin of the itinerant electrons and the localized spins on the magnetic sites $\bm{S}_i$. We emphasize that the model has no spin-orbit coupling term such that the Hamiltonian is block diagonal in spin $\sigma$.

We perform a Fourier transform and rewrite the operators in terms of electron band operators $d_{n, \bm{k}, \sigma} = \sum_{\ell} q^*_{n, \ell, \bm{k}, \sigma} c_{\ell, \bm{k}, \sigma}$. Here, $\ell$ runs over the three sites in the unit cell, $n$ runs over the energy bands, and $q^*_{n, \ell, \bm{k}, \sigma}$ is chosen such that
\begin{align}
    H_e = \sum_{n, \bm{k}, \sigma} \varepsilon_{n, \bm{k}, \sigma} d^\dagger_{n, \bm{k}, \sigma}d_{n, \bm{k}, \sigma}.
    \label{electronDispersion}
\end{align}

\begin{figure}[h!]
\includegraphics[width=0.75\columnwidth]{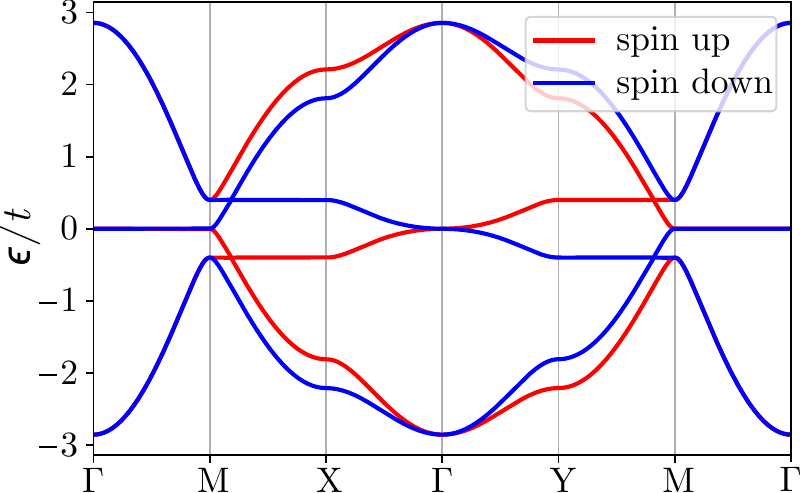}
\caption{\label{fig:electronDispersion}The electron dispersion in the $x$- and $y$-direction. We used the parameters, $t_2/t=0$, $\mu/t=0$, $\varepsilon_{\mathrm{nm}}/t = 0$ and $J_{\mathrm{sd}}S/t = 0.4$.}
\end{figure}
Fig. \ref{fig:electronDispersion} shows the electronic spectrum $\varepsilon_{n, \bm{k}, \sigma}$ of the minimal model in Eq. \eqref{electronHamiltonian}. The bands exhibit the characteristic spin splitting of altermagnets with a $d$-wave symmetry. Without next-nearest neighbor hopping, the model has an accidental particle-hole symmetry relating spin-up and spin-down at half-filling.

\begin{figure}[ht!]
\centering
\begin{subfloat}[\label{subfig:DOS1}]{\includegraphics[width=0.49\columnwidth]{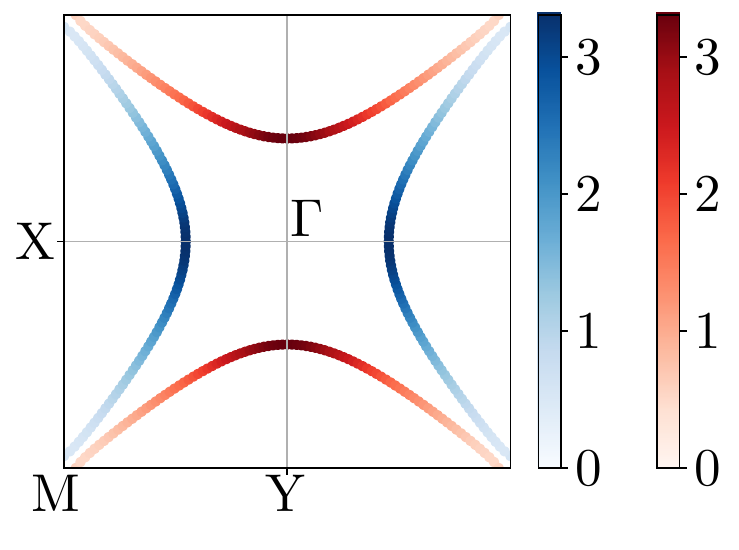}}
\end{subfloat}
\begin{subfloat}[\label{subfig:DOS2}]{\includegraphics[width=0.49\columnwidth]{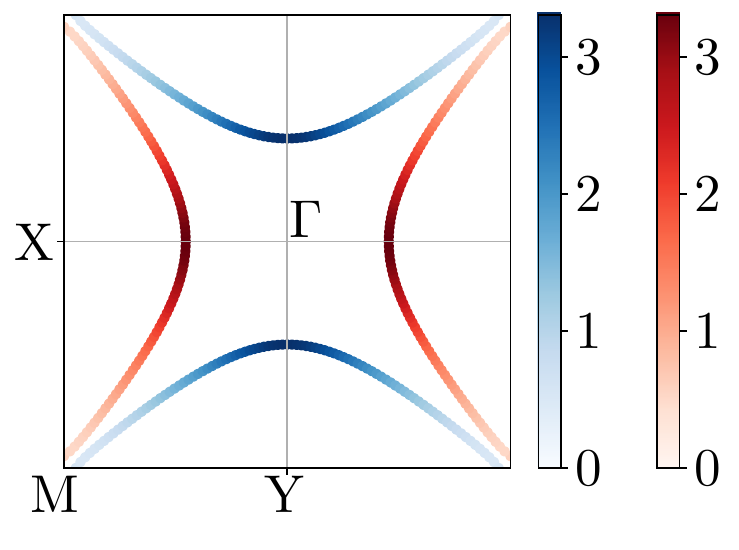}}
\end{subfloat}
\begin{subfloat}[\label{subfig:DOS3}]{\includegraphics[width=0.49\columnwidth]{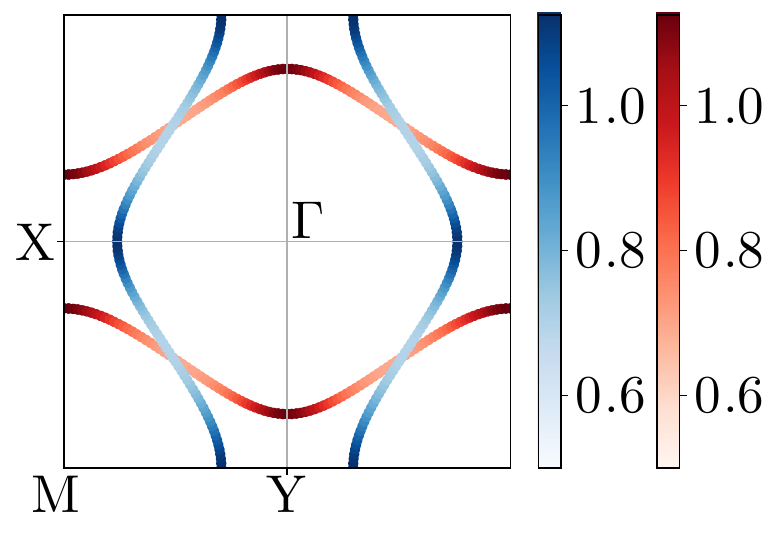}}
\end{subfloat}
\begin{subfloat}[\label{subfig:DOS4}]{\includegraphics[width=0.49\columnwidth]{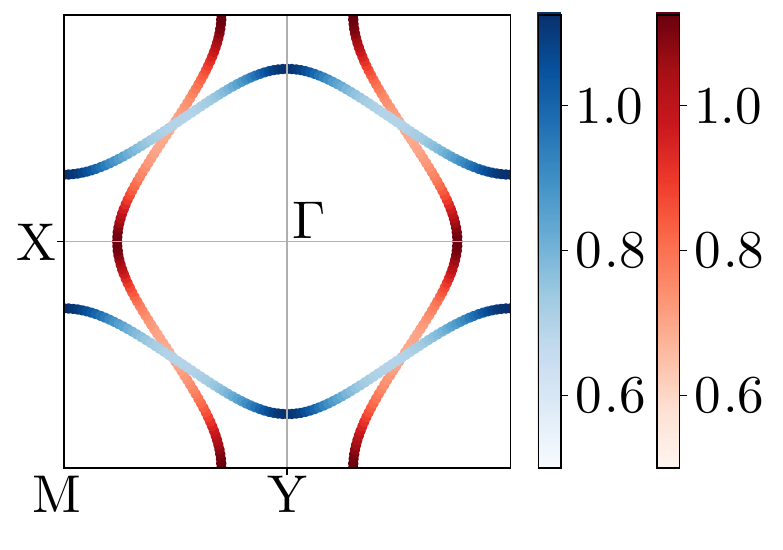}}
\end{subfloat}
\caption{\label{fig:FermiSurfaces}The Fermi surface at four distinct doping levels. Red and blue correspond to spin-up and spin-down polarization, respectively. The intensity shows the flatness of the bands $\abs{d \varepsilon/d k_{\perp}}^{-1}$, where $k_{\perp}$ is perpendicular to the Fermi surface. The plots are for $J_{\mathrm{sd}}S/t=0.4$, $t_2/t=0$, $\varepsilon_{\mathrm{nm}}/t=0$ and a) $\mu/t=0.1$, b) $\mu/t=-0.1$, c) $\mu/t=2$, d) $\mu/t=-2$.} 
\end{figure}

Fig. \ref{fig:FermiSurfaces} shows the Fermi surfaces of the electron dispersion in Fig. \ref{fig:electronDispersion} at four doping levels. The derivation is given in Appendix \ref{sec:ElectronHam}.
The mirror symmetries of the crystal in Eq. \eqref{eq:sym2} enforce the electron bands to be spin degenerate along the diagonals $k_x = \pm k_y$.
The spin-polarized sectors switch polarity as a function of the chemical potential at $\mu=0$. The exactness of the switching shown in Fig. \ref{fig:FermiSurfaces} is due to an accidental particle-hole symmetry. Nevertheless, the chemical potential $\mu$ remains an effective handle for switching the spin-polarization even without particle-hole symmetry.

\begin{figure}[ht!]
\centering
\begin{subfloat}[\label{subfig:SSm}]{\includegraphics[width=0.49\columnwidth]{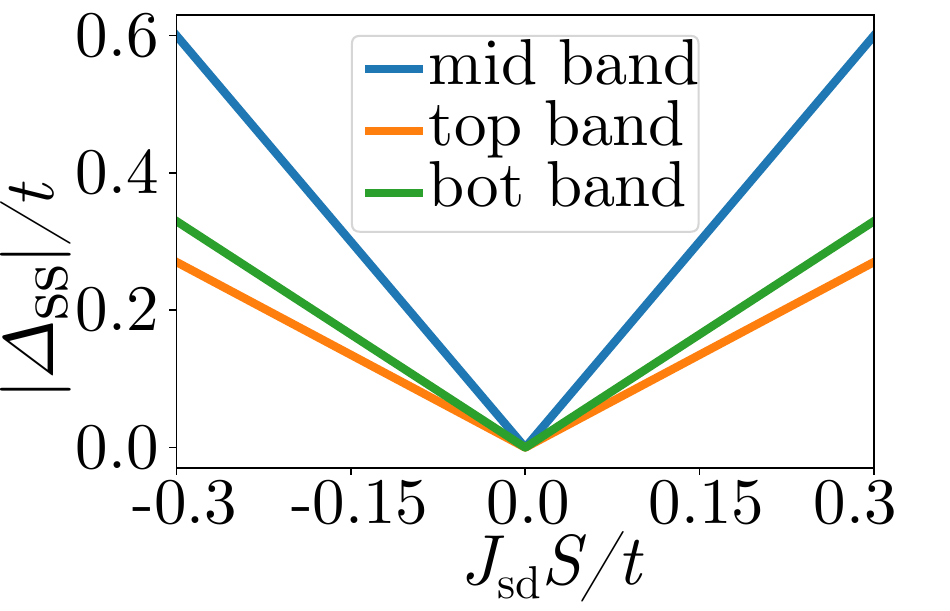}}
\end{subfloat}
\begin{subfloat}[\label{subfig:SSmu2}]{\includegraphics[width=0.49\columnwidth]{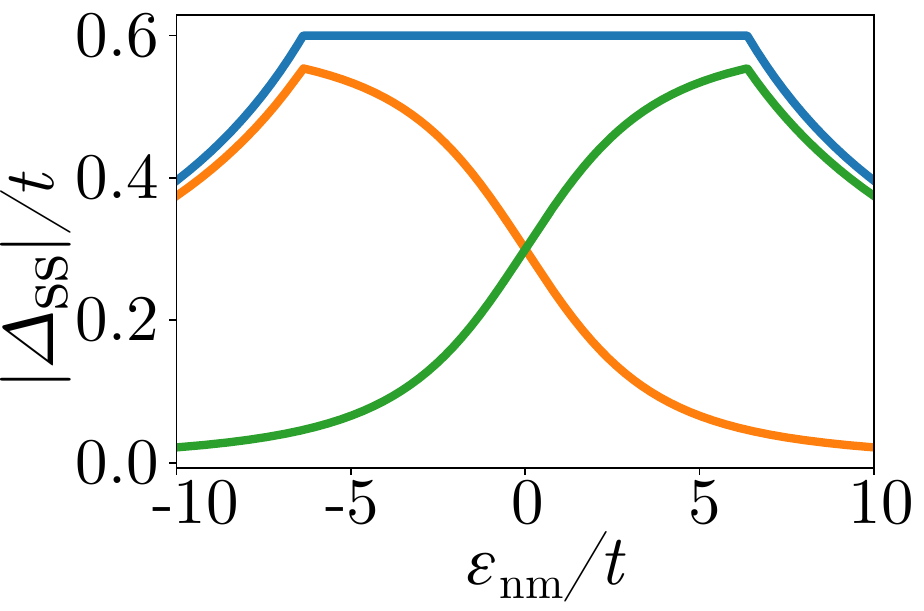}}
\end{subfloat}
\begin{subfloat}[\label{subfig:SSt}]{\includegraphics[width=0.49\columnwidth]{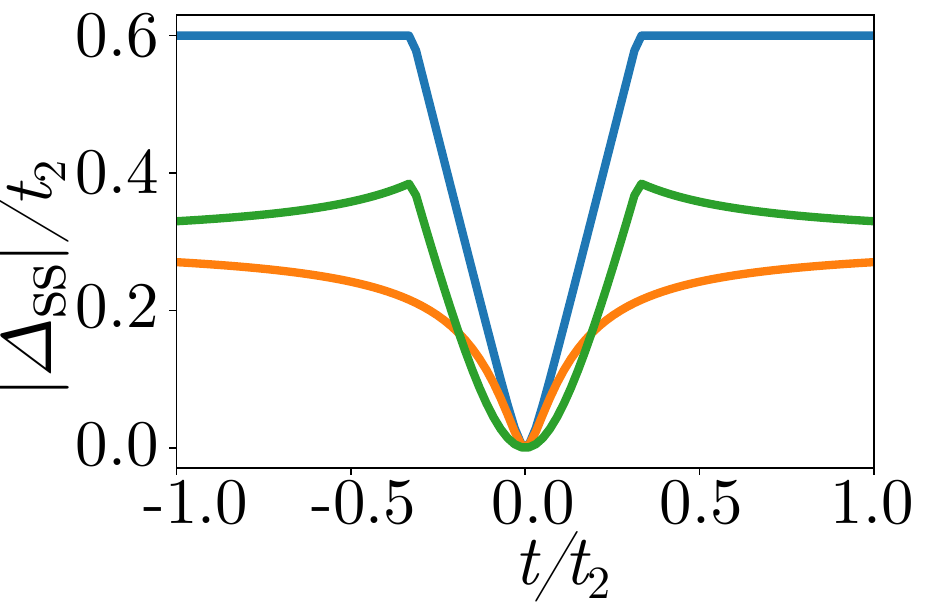}}
\end{subfloat}
\caption{\label{fig:SpinSplitting}The magnitude of the maximal spin splitting $\abs{\Delta_{\mathrm{ss}}}$ of the three bands as a function of (a) the exchange coupling $J_{\mathrm{sd}}S/t$ with $t_2/t=0$ and $\varepsilon_{\mathrm{nm}}/t = 0.4$, (b) the site-energy of the non-magnetic site $\varepsilon_{\mathrm{nm}}/t$ with $t_2/t=0$ and $J_{\mathrm{sd}}S/t=0.3$, and (c)
the nearest neighbor hopping $t/t_2$ with $\varepsilon_{\mathrm{nm}}/t_2= 0.4$ and $J_{\mathrm{sd}}S/t_2= 0.3$.} 
\end{figure}

The unconventional spin-splitting is consistent with the symmetries in Eq. \eqref{eq:whole}. However, the magnitude of the spin-splitting must be determined from microscopic calculations. Fig. \ref{fig:SpinSplitting} summarizes how spin-splitting depends on the tight-binding parameters in Eq. \eqref{electronHamiltonian}.

The spin splitting is directly proportional to the coupling strength $J_{\mathrm{sd}}S$. This is generally magnitudes larger than the spin-orbit coupling. Hence, the spin splitting in altermagnets is much larger than spin splitting due to spin-orbit coupling.
Fig. \ref{subfig:SSmu2}) shows that the site-energy $\varepsilon_{\mathrm{nm}}$ of the non-magnetic site only weakly affects the spin splitting of the middle bands. This illustrates how the presence of the non-magnetic site, but not necessarily its occupation, is essential for the altermagnetic properties.
Fig. \ref{subfig:SSt}) shows the spin splitting as a function of the nearest-neighbor hopping $t/t_2$. In the case of $t/t_2=0$, the model reduces to that of a conventional antiferromagnet with a non-magnetic site in the atomic limit. At this point, the spin splitting vanishes due to a restored $\mathcal{P}\mathcal{T}$-symmetry. 

\section{\label{sec:magnons}Magnon Properties}
Here, we consider the interactions between the localized spins shown in Fig. \ref{fig:lattice}. Nearest-neighbor interactions do not couple the magnetic sublattices. Hence, we consider the effective spin-spin exchange interactions mediated by the non-magnetic sites, the vacuum sites, and interactions along the diagonals. The inequivalence between the non-magnetic and empty sites gives rise to an anisotropic exchange interaction with a fourfold symmetry. In other words, the exchange interactions adopt the symmetry of the altermagnet. We consider
\begin{align}
\begin{split}
    H_m =& \; \sum_{\left<i,j\right>} J_{\mathrm{AB}}\bigg(\bm{S}^A_{i} \cdot \bm{S}^B_{j}\bigg) + \sum_{\left<i_x,j_x\right>} \bigg(J_{\mathrm{nm}}\bm{S}^A_{i} \cdot \bm{S}^A_{j} + J_{\mathrm{d}}\bm{S}^B_{i} \cdot \bm{S}^B_{j}\bigg) \\
    & \; + \sum_{\left<i_y,j_y\right>} \bigg(J_{\mathrm{d}}\bm{S}^A_{i} \cdot \bm{S}^A_{j} + J_{\mathrm{nm}}\bm{S}^B_{i} \cdot \bm{S}^B_{j}\bigg) \\ & \; - \sum_{i} \bigg(K_z (S^A_{z,i}S^A_{z,i} + S^B_{z,i}S^B_{z,i}) + B_z (S^A_{z,i} - S^B_{z,i})\bigg).
    \label{spinHamiltonian1}
\end{split}
\end{align}
Here, the parameters $J_{\mathrm{nm}}$ and $J_{\mathrm{d}}$ denote the strength of the exchange interactions mediated by the non-magnetic sites and the direct spin-exchange between magnetic sites absent an intermediate non-magnetic site, respectively. These terms couple equal spin sites and are thus ferromagnetic-like. Furthermore, $J_{AB}$ quantifies the interaction strength along the diagonals and couples the magnetic sublattices. It couples sites of opposite spin, giving rise to the antiferromagnetic-like exchange interaction. We also include an easy-axis anisotropy $K_z$ and magnetic field $B_z$. As outlined in Appendix \ref{sec:MagnonHam}, we treat the Hamiltonian to bilinear order in magnon operators by a Holstein-Primakoff transformation and diagonalize it with
$a_{\bm{q}} = u_{\bm{q}}\alpha_{\bm{q}} + v_{\bm{q}}\beta^\dagger_{-\bm{q}}$ and $b^\dagger_{-\bm{q}} = v^*_{\bm{q}}\alpha_{\bm{q}} + u^*_{\bm{q}}\beta^\dagger_{-\bm{q}}$. Here, $a^{(\dagger)}_{\bm{q}}$ and $b^{(\dagger)}_{\bm{q}}$ are spin (creation) and annihilation operators at the two magnetic sublattices. The magnon Hamiltonian is then given by
\begin{align}
    H_m = \sum_{\bm{q}} \bigg[\omega^\alpha_{\bm{q}} \alpha^\dagger_{\bm{q}} \alpha_{\bm{q}} + \omega^\beta_{\bm{q}} \beta^\dagger_{\bm{q}} \beta_{\bm{q}}\bigg],
\end{align}
where $\alpha_{\bm{q}}^{(\dagger)}$ and $\beta_{\bm{q}}^{(\dagger)}$ are the (creation) annihilation operators of the magnons with momentum $\bm{q}$. The magnon frequencies are $\omega^{\alpha(\beta)}_{\bm{q}} = \varpm \gamma_1 + \sqrt{\gamma_2^2-\gamma_3^2}$ with
\begin{subequations}
\begin{align}
    \gamma_1 =&\; S(J_{\mathrm{nm}}-J_{\mathrm{d}})(\cos{(2q_x a)} - \cos{(2q_y a)}) + B_z, \\
    \begin{split}
    \gamma_2 = &\; S(J_{\mathrm{nm}}+J_{\mathrm{d}})(\cos{(2q_x a)} + \cos{(2q_y a)}) \\ & \;- 2S(J_{\mathrm{nm}} + J_{\mathrm{d}}) +  4SJ_{AB} + 2SK_z,
    \end{split}\\
    \gamma_3 = &\; 2J_{AB} S (\cos{(q_x a + q_y a)} + \cos{(q_x a - q_y a)}).
\end{align}
\end{subequations}

\begin{figure}[h!]
\includegraphics[width=0.75\columnwidth]{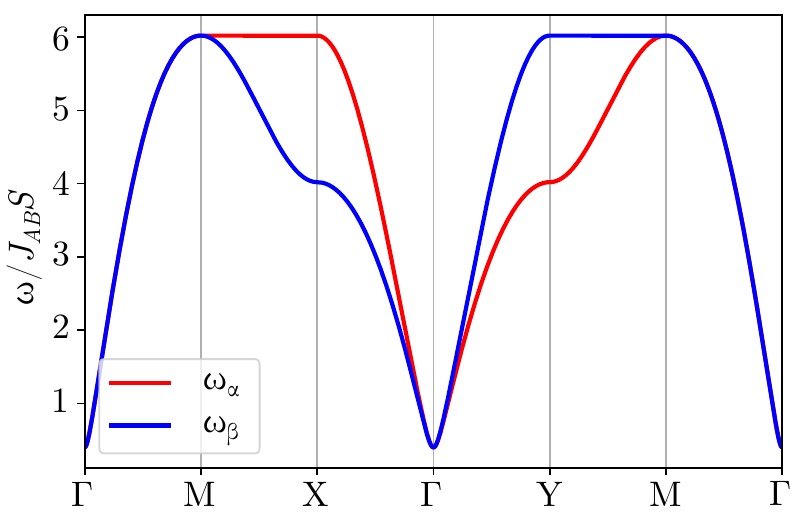}
\caption{\label{fig:MagnonDispersion}The magnon dispersion $\omega/J_{AB} S$. Here, we used the parameters $J_{\mathrm{nm}}/J_{AB} = -0.5$, $J_{\mathrm{d}}/J_{AB} = 0 $, $K_z/J_{AB} = 0.01$ and $B_z = 0$.}
\end{figure}
The magnon bands in Fig. \ref{fig:MagnonDispersion} are split and carry a finite spin expectation value. However, they are not isolated and thus not topological. The mirror symmetry forces band degeneracies along the diagonal without an external magnetic field.

The magnon-band splitting has a $d$-wave symmetry in momentum space given by
\begin{align}
    \Delta_{\mathrm{magnon}} = 2\left[S(J_{\mathrm{nm}}-J_{\mathrm{d}})\bigg(\cos{(2q_x a)}-\cos{(2q_y a)}\bigg)  + B_z\right].
    \label{magnonGap}
\end{align}
In other words, the magnon-band splitting arises from inequivalence between the two ferromagnetic-like exchange interaction strengths $J_{\mathrm{nm}}$ and $J_{\mathrm{d}}$.

\section{\label{sec:electronmagnon}Electron-magnon coupling}
Fluctuations in the localized spins $\bm{S}_i$ induce an effective electron-magnon coupling from the third term in Eq. \eqref{electronHamiltonian}. We treat this term using the Holstein-Primakoff transformation to second order in magnon operators. The resulting electron-magnon coupling consists of two contributions. The first term is spin-flip processes from first-order magnon operators. This term is generally important for magnon-mediated superconductivity \cite{RohlingPRB2018, FjarbuPRB2019}. However, for altermagnets, they do not contribute to robust superconducting instabilities.

The second part consists of terms that are second order in magnon operators
\begin{align}
\begin{split}
    H_{\mathrm{em}} = \frac{J_{\mathrm{sd}}}{N}\sum_{\bm{k},\bm{q}, \bm{q}', \sigma}\bigg( \sigma M^B_{\bm{q}, \bm{q}'}( \Omega^B_{\bm{k} + \bm{q} -\bm{q}', \bm{k},\sigma, \sigma} d^\dagger_{\bm{k}+\bm{q} -\bm{q}',\sigma}d_{\bm{k},\sigma}) \\ - \sigma M^A_{\bm{q}, \bm{q}'}
    (\Omega^A_{\bm{k} + \bm{q} -\bm{q}',\bm{k},\sigma, \sigma}d^\dagger_{\bm{k}+\bm{q} -\bm{q}',\sigma}d_{\bm{k},\sigma} )\bigg).
    \label{secondOrderMagnonSD}
\end{split}
\end{align}
Here, $N$ is the number of unit cells. We have defined $\Omega^{A}_{\bm{k}',\bm{k},\sigma', \sigma} \equiv q^*_{a, \bm{k}', \sigma'}q_{a, \bm{k}, \sigma}$ and $\Omega^{B}_{\bm{k}',\bm{k},\sigma', \sigma} \equiv q^*_{b, \bm{k}', \sigma'}q_{b, \bm{k}, \sigma}$, where we dropped the band index $n$ because we consider scattering processes on the Fermi surface. The magnon terms are
\begin{subequations}
\begin{align}
    \begin{split}
    M^A_{\bm{q}, \bm{q}'} = u^*_{\bm{q}'} u_{\bm{q}} \alpha^\dagger_{\bm{q}'} \alpha_{\bm{q}} + u^*_{\bm{q}'} v_{\bm{q}} \alpha^\dagger_{\bm{q}'}\beta^\dagger_{-{\bm{q}}} \\+ v^*_{\bm{q}'}u_{\bm{q}} \beta_{-\bm{q}'}\alpha_{\bm{q}} + v^*_{\bm{q}'} v_{\bm{q}}\beta_{-\bm{q}'} \beta^\dagger_{-\bm{q}}, 
    \end{split}
    \\
    \begin{split}
    M^B_{\bm{q}, \bm{q}'} = v^*_{\bm{q}'}v_{\bm{q}} \alpha_{-\bm{q}'} \alpha^\dagger_{-\bm{q}} + v^*_{\bm{q}'} u_{\bm{q}} \alpha_{-\bm{q}'}\beta_{\bm{q}} \\+ u^*_{\bm{q}'}v_{\bm{q}}\beta_{\bm{q}'}^\dagger \alpha^\dagger_{-\bm{q}} + u^*_{\bm{q}'} u_{\bm{q}}\beta^\dagger_{\bm{q}'}\beta_{\bm{q}}.
    \end{split}
\end{align}
\end{subequations}

Based on the electron-magnon coupling, we derive an effective electron-electron interaction using a Schrieffer-Wolff transformation \cite{SchriefferWolff1966}. Finite momentum Cooper pairs are frail. Hence, we restrict our considerations to interactions between electrons of opposite momenta.
Consequently, spin-flip processes associated with the first-order magnon terms are prohibited, except at the spin-degenerate band crossings. These crossings are extremely material-dependent and tend to be point-like at the Fermi surface. We rule them out as secondary effects in the superconducting properties of altermagnets and do not explore this interaction further.

As shown in Appendix \ref{sec:EffectiveInt}, the spin-conserving scattering processes give rise to an effective electron-electron interaction
\begin{align}
    H_{\mathrm{e-e}} = \sum_{\bm{k}, \bm{k}', \sigma} V_{\bm{k}, \bm{k}', \sigma} d^\dagger_{\bm{k}',\sigma}d^\dagger_{-\bm{k}',\sigma}d_{-\bm{k},\sigma} d_{\bm{k},\sigma},
\end{align}
where
\begin{align}
\begin{split}
    V_{\bm{k}, \bm{k}', \sigma} = \frac{-J^2_{\mathrm{sd}}}{N^2}\sum_{\bm{Q}} \bigg(\Omega^A_{\bm{k}',\bm{k},\sigma, \sigma}
     \abs{v_{\frac{\bm{Q}-(\bm{k}'-\bm{k})}{2}}} \abs{u_{\frac{\bm{Q}+(\bm{k}'-\bm{k})}{2}}} \\- \Omega^B_{\bm{k}',\bm{k},\sigma, \sigma}
     \abs{v_{\frac{\bm{Q}+(\bm{k}'-\bm{k})}{2}}} \abs{u_{\frac{\bm{Q}-(\bm{k}'-\bm{k})}{2}}}\bigg)^2 \\ \times \bigg(\omega^\alpha_{\frac{\bm{Q}+(\bm{k}'-\bm{k})}{2}} + \omega^\beta_{\frac{\bm{Q}-(\bm{k}'-\bm{k})}{2}}\bigg)^{-1}.
    \label{V2}
\end{split}
\end{align}
Here, we sum over the total magnon momentum $\bm{Q}=\bm{q}+\bm{q}'$.
For equal spin Cooper pairs, the interaction has to be odd in momenta $\bm{k}$ and $\bm{k}'$ to respect the Pauli principle. We symmetrize the effective interaction and keep the odd contribution.

\begin{figure}[ht!]
\centering
\begin{subfloat}[\label{subfig:GapProfileLarge}]{\includegraphics[width=0.49\columnwidth]{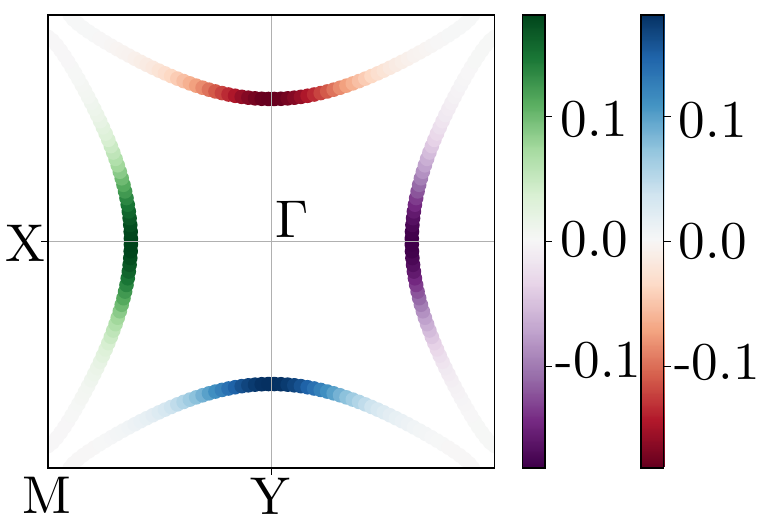}}
\end{subfloat}
\begin{subfloat}[\label{subfig:GapProfileLow}]{\includegraphics[width=0.49\columnwidth]{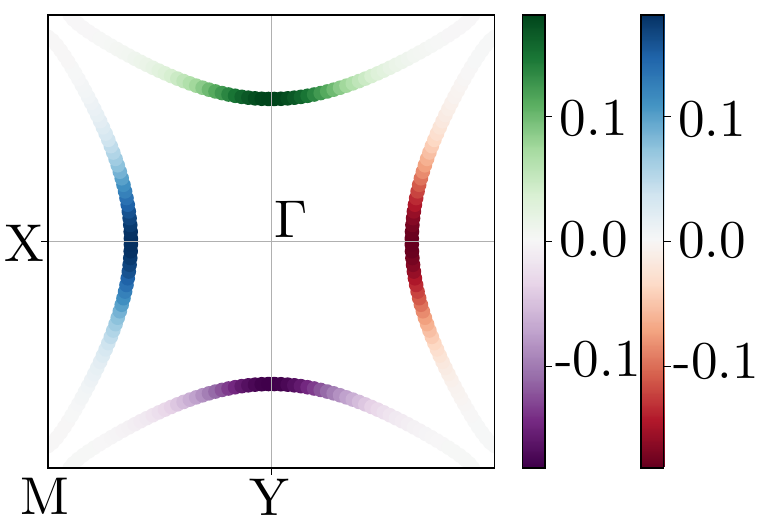}}
\end{subfloat}
\caption{\label{fig:GapProfiles}The superconducting gap profiles of $\Delta_{\bm{k}, \sigma}$ corresponding to the superconducting state with the critical temperature $T_c$ on the Fermi surface. Red-blue denotes the gap $\Delta_{\bm{k}, \uparrow}$ and green-purple denotes $\Delta_{\bm{k}, \downarrow}$. Both gaps are odd in momentum $\bm{k}$ and correspond to $p$-wave superconductivity. The parameters are $J_{\mathrm{sd}}S/t=0.4$ and a) $\mu/t=0.2$,  b) $\mu/t=-0.2$. The magnon dispersion has a smaller energy scale with $J_{AB}S/t = 0.01$. We set $J_{\mathrm{nm}}/J_{AB}=J_{\mathrm{d}}/J_{AB}=-0.5$ and $K_z/J_{AB} = 0.01$.}
\end{figure}

The superconducting gap is $\Delta_{\bm{k},\sigma} = -\sum_{\bm{k'}}V_{\bm{k},\bm{k}',\sigma}\left<d_{-\bm{k}',\sigma}  d_{\bm{k}',\sigma}\right>$,
and gives the gap equation
\begin{align}
\begin{split}
    \Delta_{\bm{k},\sigma}
    = -\sum_{\bm{k}'}V_{\bm{k},\bm{k}',\sigma}\frac{\Delta_{\bm{k}',\sigma}}{E_{\bm{k}',\sigma}}\tanh{\left(\frac{\beta E_{\bm{k}',\sigma}}{2}\right)},
    \label{GapEquation}
\end{split}
\end{align}
with a quasiparticle dispersion
$E_{\bm{k},\sigma} = \sqrt{\varepsilon_{\bm{k},\sigma}^2 + 4\abs{\Delta_{\bm{k},\sigma}}^2}$.
We linearize the gap equation and perform a Fermi surface average. We take the resulting eigenvalue equation and solve for the largest eigenvalue $\lambda_{\mathrm{eff}}$. The well-known BCS formula \cite{BardeenPR1957} gives the corresponding critical temperature
\begin{align}
    T_c \approx \frac{1.13\omega_M}{k_B} e^{-\frac{1}{\lambda_{\mathrm{eff}}}}.
    \label{CriticalTemperature}
\end{align}
The $\bm{k}$-dependent gap profile $\Delta_{\bm{k}, \sigma}$ is given by the corresponding eigenvector. Fig. \ref{fig:GapProfiles} shows the gap profile for two distinct values of $\mu$. We note how the doping changes the spin-polarization of the bands and serves as a spin switch.
The quasiparticle dispersion is fully gapped because the Fermi surface is absent at the nodes. Hence, we expect superconductivity in altermagnets to be robust against weak magnetic- and non-magnetic disorder. For triplet superconductors, the order parameter can be parametrized by the complex $\bm{d}$-vector \cite{SigristUeda}. It is defined as
\begin{align}
\begin{split}
    \hat{\Delta}(\bm{k}) =& \begin{pmatrix} \Delta_{\uparrow \uparrow}(\bm{k}) && \Delta_{\uparrow \downarrow}(\bm{k}) \\
    \Delta_{\downarrow \uparrow}(\bm{k}) && \Delta_{\downarrow \downarrow}(\bm{k})\end{pmatrix} \\=& \begin{pmatrix} -d_x + id_y && d_z \\
    d_z && d_x + id_y\end{pmatrix} = i(\bm{d}(\bm{k})\cdot\bm{\sigma})\sigma_y.
\end{split}
\end{align}
To parameterize the gap profile in Fig. \ref{subfig:GapProfileLarge}, we use two $\bm{d}$-vectors to capture the gap profiles of the disjoint spin-polarized bands. The gap is parametrized by
\begin{subequations}
\begin{align}
    \bm{d}_1 =& \; \frac{\Delta(\abs{\bm{k}})}{\sqrt{2}}\sign(k_x)(-1, i, 0), \\
    \bm{d}_2 =& \; \frac{\Delta(\abs{\bm{k}})}{\sqrt{2}}\sign(k_y)(1, i, 0).
\end{align}
\end{subequations}
Here, the subscripts $1$ and $2$ correspond to the spin-down and spin-up bands, respectively.
The corresponding spin-polarization of the superconducting order parameter for the two bands are
\begin{subequations}
\begin{align}
    \bm{q}_1 =& \; i(\bm{d}_1 \times \bm{d}^*_1) = \big(0, 0, -\abs{\Delta(\abs{\bm{k}})}^2\big), \\
    \bm{q}_2 =& \; i(\bm{d}_2 \times \bm{d}^*_2) = \big(0, 0, \abs{\Delta(\abs{\bm{k}})}^2\big),
\end{align}
\end{subequations}
consistent with the bands being fully spin-polarized.

\begin{figure}[ht!]
\centering
\begin{subfloat}[\label{subfig:lambdas}]{\includegraphics[width=0.49\columnwidth]{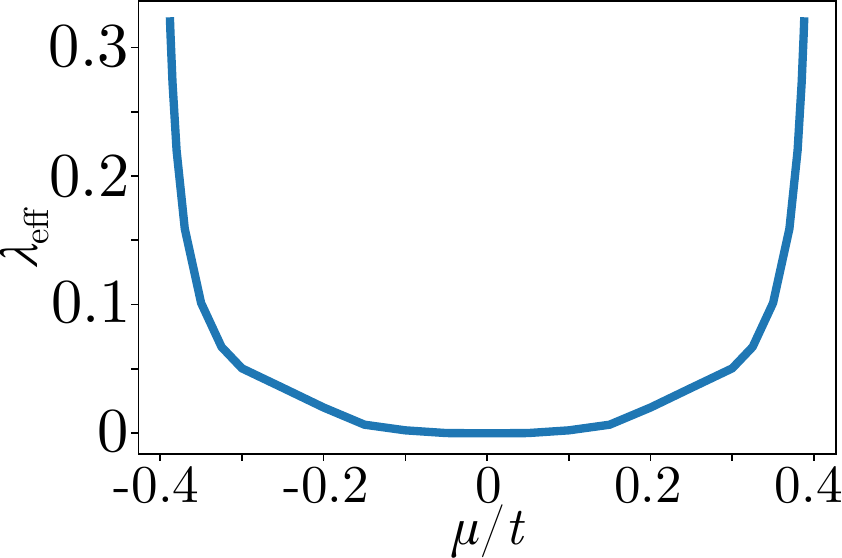}}
\end{subfloat}
\begin{subfloat}[\label{subfig:Tc}]{\includegraphics[width=0.49\columnwidth]{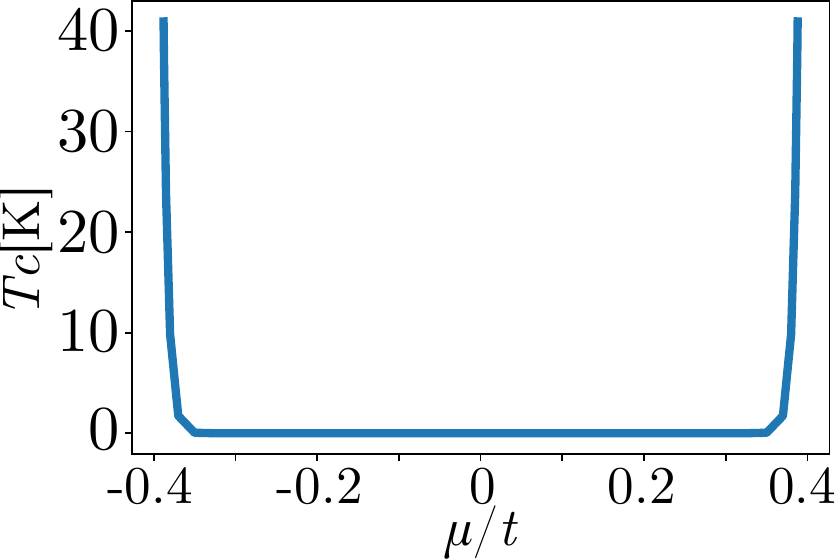}}
\end{subfloat}
\begin{subfloat}[\label{subfig:Omegas}]{\includegraphics[width=0.49\columnwidth]{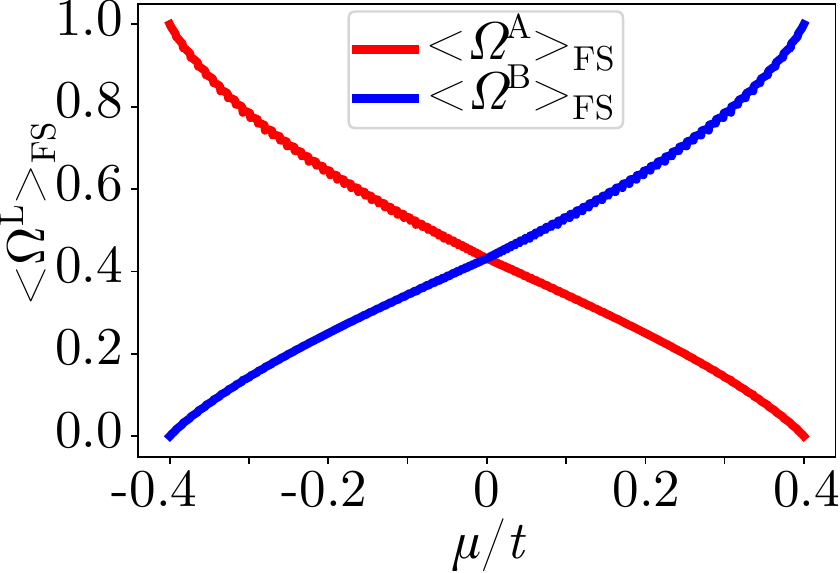}}
\end{subfloat}
\caption{\label{fig:Squeezing}a) The interaction strength, b) critical temperature, and c) $\left<\Omega^A_{\bm{k}, \bm{k},\uparrow, \uparrow}\right>_{\mathrm{FS}}$ and $\left<\Omega^B_{\bm{k}, \bm{k},\uparrow, \uparrow}\right>_{\mathrm{FS}}$ as functions of the chemical potential $\mu$. Here, $J_{\mathrm{sd}}S/t= 0.4$, and $t_2/t = 0$. Furthermore, $J_{AB}S/t = 0.01$, $K_z/J_{AB} = 0.01$ and the magnitude of the localized spins $S=3/2$. The ferromagnetic-like exchange couplings are $J_{\mathrm{nm}}/J_{AB}=J_{\mathrm{d}}/J_{AB}=-0.5$.} 
\end{figure}

The estimate of the critical temperature in Eq. \eqref{CriticalTemperature} is notoriously unreliable due to its exponential sensitivity to material properties. Nevertheless, it could give an idea of the order of magnitude and the qualitative behavior of the superconducting state. Fig. \ref{fig:Squeezing} highlights the qualitative relation between the occupation of the magnetic sites and the critical temperature $T_c$. The figure shows a dramatically enhanced critical temperature for certain values of $\mu$. This can be understood as an intrinsic analog to the squeezing effect in Refs. \cite{Erlandsen2019PRB, Erlandsen2021PRB}, without an uncompensated interface.

\section{\label{sec:Conclusion}Conclusion}
We have introduced a minimal model for altermagnets. The model exhibits the expected behaviors for an altermagnet and allows for tuning the altermagnetic properties through simple tight-binding parameters. We explore the electron- and magnon properties of the model. The electron-magnon coupling gives rise to spin-polarized $p$-wave superconducting states. Furthermore, we find that the critical temperature can be significantly enhanced by tuning the chemical potential. This effect is an intrinsic analog to the magnon-squeezing effect predicted for antiferromagnets with uncompensated interfaces.

\textit{While finalizing the manuscript, we found a study that considers a similar magnon model. They consider thermal magnon transport} \cite{cui2023efficient}.

\section{\label{sec:Acknowledgements}Acknowledgements}
We thank Kristian Mæland for useful feedback on the manuscript. The Research Council of Norway (RCN) supported this work through its Centres of Excellence funding scheme, project number 262633, "QuSpin", and RCN project number 323766.

\appendix

\section{\label{sec:ElectronHam}Electron Hamiltonian}

We consider the altermagnetic square lattice in Fig. \ref{fig:lattice}.
The lattice vectors  are
\begin{align}
    \bm{a}_1 = 2a(1,0), && \bm{a}_2 = 2a(0,1),
\end{align}
where the nearest-neighbor distance $a=1$.
The reciprocal lattice vectors are
\begin{align}
    \bm{b}_1 = \frac{\pi}{a}(1,0), && \bm{b}_2 = \frac{\pi}{a}(0,1).
\end{align}

We Fourier transform the electron operators in Eq. \eqref{electronHamiltonian} as
\begin{align}
    c_{\ell, i,\sigma}= \frac{1}{\sqrt{N}}\sum_{\bm{k}} c_{\ell, \bm{k},\sigma} e^{-i\bm{k}\cdot\bm{r}_i},
    \label{FourierTransform}
\end{align}
where we assume periodic boundary conditions in both the $x$- and $y$-direction with a total of $N$ unit cells. The crystal momentum $\bm{k}$ runs over the first Brillouin zone. The subscript $\ell$ runs over the three sites in the unit cell $\ell \in (a, \mathrm{nm}, b)$, where $a$ and $b$ refer to the magnetic sites and $\mathrm{nm}$ refers to the non-magnetic site.
In this basis, the Hamiltonian is
\begin{align}
    H_e = \sum_{\bm{k}, \sigma, \ell, \ell'}c_{\ell,\bm{k},\sigma}^\dagger
    \big(\mathcal{H}_{\bm{k}, \sigma}\big)_{\ell, \ell'}c_{\ell',\bm{k},\sigma},
\end{align}
where
\begin{widetext}
\begin{align}
    \mathcal{H}_{\bm{k}, \sigma} = \begin{pmatrix} -\sigma J_{\mathrm{sd}} S-\mu && 2t\cos{k_x a} && 2 t_2\cos{(k_x a +k_y a)} + 2 t_2\cos{(k_x a-k_y a)} \\ 2t\cos{k_x a} && -\mu + \varepsilon_{\mathrm{nm}} && 2t\cos{k_y a} \\ 2 t_2\cos{(k_x a +k_y a)} + 2 t_2\cos{(k_x a-k_y a)} && 2t\cos{k_y a} && \sigma J_{\mathrm{sd}} S -\mu
    \end{pmatrix}.
    \label{3x3Matrix}
\end{align}
\end{widetext}

We diagonalize this Hamiltonian by introducing the electron band operators
\begin{align}
    d_{n,\bm{k},\sigma} =  \sum_{\ell} q^*_{n, \ell, \bm{k}, \sigma} c_{\ell, \bm{k}, \sigma}, && d^\dagger_{n,\bm{k},\sigma} = \sum_{\ell} q_{n, \ell, \bm{k}, \sigma} c^\dagger_{\ell,\bm{k},\sigma} .
\end{align}
The diagonal form of the electron Hamiltonian is given in Eq. \eqref{electronDispersion}.

\subsection{The Fermi surface}
The electron band energies in Eq. \eqref{electronDispersion} are solutions to the secular equation
\begin{align}
\begin{split}
    x_0 + x_1 \varepsilon + x_2 \varepsilon^2 + \varepsilon^3 = 0.
    \label{secularEq}
\end{split}
\end{align}
\begin{widetext}
where the coefficients are
\begin{subequations}
\begin{align}
\begin{split}
    x_0 = \mu^3 - \mu J^2_{\mathrm{sd}} S^2 - \mu^2 \varepsilon_{\mathrm{nm}} + J^2_{\mathrm{sd}} S^2 \varepsilon_{\mathrm{nm}} - 4 \mu t^2 - 8 t^2 t_2 + 
 4 (-\mu + \varepsilon_{\mathrm{nm}}) t_2^2 + \\
 2 ((-\mu -\sigma J_{\mathrm{sd}} S) t^2 - 4 t^2 t_2 + 2 (-\mu + \varepsilon_{\mathrm{nm}}) t_2^2) \cos{(2 k_y a)} + 
 2 \cos{(2 k_x a)} \big((-\mu +\sigma J_{\mathrm{sd}} S) t^2 - 4 t^2 t_2 + 2 (-\mu + \varepsilon_{\mathrm{nm}}) t_2^2 \\+ 
    2 t_2 (-2 t^2 + (-\mu + \varepsilon_{\mathrm{nm}}) t_2) \cos{(2 k_y a)}\big),
\end{split}
\end{align}

\begin{align}
\begin{split}
    x_1 = 3 \mu^2 - J^2_{\mathrm{sd}} S^2 - 2 \mu\varepsilon_{\mathrm{nm}} - 4 (t^2 + t_2^2) - 
     2 t^2 \cos{(2 k_x a)} - 4 t_2^2 \cos{(2 k_x a)} - 2 t_2^2 \cos{(2 k_x a - 2 k_y a)} \\- 
     2 t^2 \cos{(2 k_y a)} - 4 t_2^2 \cos{(2 k_y a)} - 
     2 t_2^2 \cos{(2 k_x a + 2 k_y a)},
\end{split}
\end{align}
\begin{align}
     x_2 = 3 \mu -\varepsilon_{\mathrm{nm}}.
\end{align}
\end{subequations}
\end{widetext}
The Fermi surface is the contour defined by $\varepsilon=0$ such that Eq. \eqref{secularEq} reduces to $x_0 = 0$. We solve for $k_x$ and $k_y$ explicitly to find
\begin{subequations}
\begin{align}
    k_y = \frac{1}{2a}\arccos{\bigg(\frac{-\alpha -\gamma \cos{(2k_x a)}}{\beta+\delta\cos{(2k_x a)}}\bigg)}
    \label{FermiSurface}
\end{align}
and conversely
\begin{align}
    k_x = \frac{1}{2a}\arccos{\bigg(\frac{-\alpha -\beta \cos{(2k_y a)}}{\gamma+\delta\cos{(2k_y a)}}\bigg)}.
\end{align}
\end{subequations}
Here,
\begin{subequations}
\begin{align}
    \begin{split}
    \alpha =&\; \mu^3 - \mu J^2_{\mathrm{sd}} S^2 - \mu^2 \varepsilon_{\mathrm{nm}} + J^2_{\mathrm{sd}} S^2 \varepsilon_{\mathrm{nm}} \\ &- 4 \mu t^2 - 8 t^2 t_2 + 4 (-\mu + \varepsilon_{\mathrm{nm}}) t_2^2,
    \end{split}
    \\
    \beta =&\; 2 ((-\mu -\sigma J_{\mathrm{sd}}S) t^2 - 4 t^2 t_2 + 2 (-\mu + \varepsilon_{\mathrm{nm}}) t_2^2),
    \\
    \gamma =&\; 2 ((-\mu + \sigma J_{\mathrm{sd}}S) t^2 - 4 t^2 t_2 + 2 (-\mu + \varepsilon_{\mathrm{nm}}) t_2^2 ),
    \\
    \delta =&\; 4 t_2 (-2 t^2 + (-\mu + \varepsilon_{\mathrm{nm}}) t_2).
\end{align}
\end{subequations}

\subsection{Sample the Fermi surface}
We solve the superconducting gap equation \eqref{GapEquation} by employing a Fermi surface average. This requires a uniform sampling of the Fermi surface. To that end, we parametrize the Fermi surface in Eq. \eqref{FermiSurface} as
\begin{align}
    k_x = t, && k_y = \frac{1}{2a}\arccos{\bigg(\frac{-\alpha -\gamma \cos{(2ta)}}{\beta+\delta\cos{(2 t a)}}\bigg)},
    \label{parametrization}
\end{align}
where $t \in [-\pi/2a, \pi/2a]$.
For a fixed segment $\Delta t$ along the $k_x$-axis, the corresponding segment along the Fermi surface is
\begin{align}
    \Delta s = \Delta t \sqrt{\bigg(\frac{d k_x}{d t}\bigg)^2 + \bigg(\frac{d k_y}{d t}\bigg)^2},
\end{align}
where
\begin{align}
    \frac{d k_y}{d t} = -\frac{1}{\sqrt{1-\bigg(\frac{\alpha+\gamma\cos{(2ta)}}{\beta + \delta\cos{(2ta)}}\bigg)^2}}\frac{\big(\gamma\beta - \delta \alpha\big)\sin{(2ta)}}{\big(\beta + \delta\cos{(2ta)}\big)^2}.
\end{align}
Now, setting $\Delta s$ constant and sample Eq. \eqref{parametrization} at intervals
\begin{align}
    \Delta t =  \frac{\Delta s}{\sqrt{\bigg(1 + \big(\frac{d k_y}{d t}\big)^2\bigg)}}
    \label{sampleInterval}
\end{align}
gives a uniform distribution along the Fermi surface.

\section{\label{sec:MagnonHam}Magnon Hamiltonian}

\begin{figure}[t!]
\centering
\begin{subfloat}[\label{subfig:}]{\includegraphics[width=0.4\columnwidth]{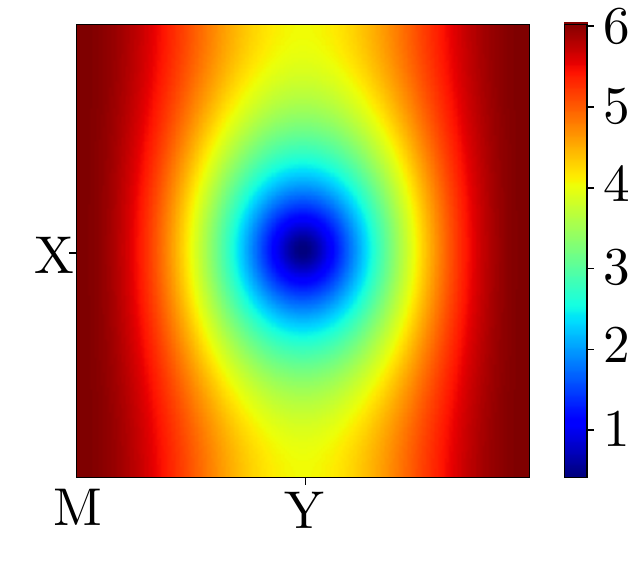}}
\end{subfloat}
\begin{subfloat}[\label{subfig:FSNoDoping}]{\includegraphics[width=0.4\columnwidth]{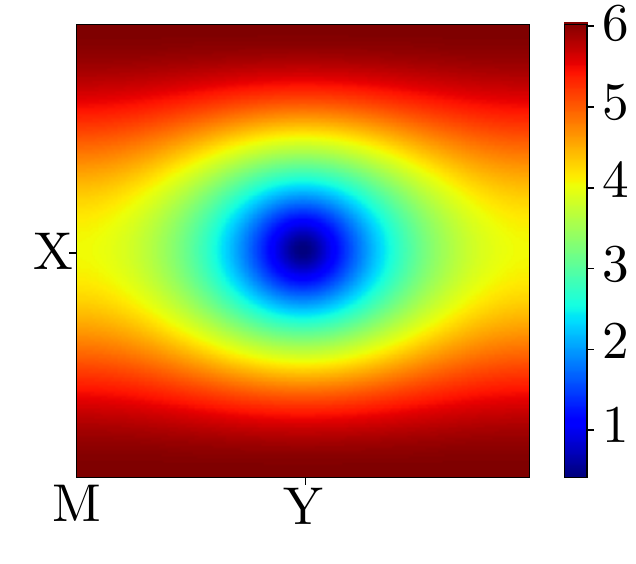}}
\end{subfloat}
\caption{\label{fig:MagnonDispersionSupp}The magnon modes with $J_{\mathrm{nm}}/J_{AB} = -0.5$, $J_{\mathrm{d}}/J_{AB} = 0$, and $K_z/J_{d} = 0.01$. Here, a) shows $\omega^{\alpha}_{\bm{q}}/J_{\mathrm{AB}}S$ and b) shows $\omega^{\beta}_{\bm{q}}/J_{\mathrm{AB}}S$.} 
\end{figure}

We consider the Hamiltonian in Eq. \eqref{spinHamiltonian1} and rewrite the spin operators as 
\begin{subequations}
\begin{align}
    S_i^{A,x} = \frac{1}{2}(S_i^{A,+} + S_i^{A,-}), && 
    S_i^{A,y} = \frac{1}{2i}(S_i^{A,+} - S_i^{A,-}), \\ S_j^{B,x} = \frac{1}{2}(S_j^{B,+} + S_j^{B,-}), && 
    S_j^{B,y} = \frac{1}{2i}(S_j^{B,+} - S_j^{B,-}).
\end{align}
\end{subequations}
Then, we quantize the spin operators to linear order using
\begin{subequations}
\begin{align}
    S_i^{A,+}=\sqrt{2S}a_i, && S_j^{B,+}=\sqrt{2S}b_j^\dagger, \\ S_i^{A,-}=\sqrt{2S}a_i^\dagger, && S_j^{B,-}=\sqrt{2S}b_j, \\ S_i^{A,z} = (S -a_i^\dagger a_i), && S_j^{B,z} = (-S + b_j^\dagger b_j),
\end{align}
\label{HolsteinPrimakoff}
\end{subequations}
where $S$ is the magnitude of the localized spins.
To diagonalize the Hamiltonian, we use the Fourier transform
\begin{align}
    a_i = \frac{1}{\sqrt{N}}\sum_{\bm{q}} a_{\bm{q}} e^{-i\bm{q}\cdot\bm{r}_i},
    \label{FourierTransformMagnon}
\end{align}
with the same convention for the $b$-operator.
We use the Holstein-Primakoff transformation and subsequent Fourier transform to find
\begin{widetext}
\begin{align}
\begin{split}
    H_m = \sum_{\bm{q}} \left[ S\bigg(2J_{\mathrm{nm}}\cos{(2q_x a)} + 2J_{\mathrm{d}}\cos{(2q_y a)} -2(J_{\mathrm{d}}+J_{\mathrm{nm}})\bigg) +  4S J_{AB} + 2 S K_z + B_z 
 \right]a_{\bm{q}}^\dagger a_{\bm{q}}\\ +\left[ S\bigg(2J_{\mathrm{nm}}\cos{(2q_y a)} + 2J_{\mathrm{d}}\cos{(2q_x a)} -2(J_{\mathrm{d}}+J_{\mathrm{nm}})\bigg) +  4S J_{AB} + 2 S K_z - B_z 
 \right]b_{\bm{q}}^\dagger b_{\bm{q}} \\
 +\left[2J_{AB} S\bigg(\cos{(q_x a+ q_y a)}+\cos{(q_x a - q_y a)}\bigg)\right]a_{\bm{q}}b_{-\bm{q}}
 \\
 +\left[ 2J_{AB} S \bigg(\cos{(q_x a+ q_y a)}+\cos{(q_x a - q_y a)}\bigg)\right]a^\dagger_{\bm{q}}b^\dagger_{-\bm{q}}.
    \label{CleanFourierMagnonHamiltonian2}
\end{split}
\end{align}
\end{widetext}
In a matrix form, the Hamiltonian is
\begin{align}
    H_{m} = \sum_{\bm{q}} \begin{pmatrix} a_{\bm{q}}^\dagger & b_{-\bm{q}} \end{pmatrix}\begin{pmatrix}A(\bm{q}) && B(\bm{q}) \\ B(\bm{q}) && C(\bm{q})\end{pmatrix}\begin{pmatrix} a_{\bm{q}} \\ b_{-\bm{q}}^\dagger \end{pmatrix},
\end{align}
where $A,B,C$ can be read off from Eq. \ref{CleanFourierMagnonHamiltonian2}.
We diagonalize the Hamiltonian by introducing the Bogoliubov transformation
\begin{align}
    \begin{pmatrix} a_{\bm{q}} \\ b_{-\bm{q}}^{\dagger}\end{pmatrix} = \begin{pmatrix} u_{\bm{q}} & v_{\bm{q}} \\ v_{\bm{q}}^* & u^*_{\bm{q}}\end{pmatrix} \begin{pmatrix} \alpha_{\bm{q}} \\ \beta_{-\bm{q}}^\dagger\end{pmatrix}.
\end{align}
The transformation must respect bosonic commutation relations, which leads to the constraint
\begin{align}
    \abs{u_{\bm{q}}}^2 - \abs{v_{\bm{q}}}^2 = 1.
    \label{commutation}
\end{align}
We choose
\begin{subequations}
\begin{align}
    u_{\bm{q}} = \frac{i}{\sqrt{2}}\sqrt{\frac{A+C}{\sqrt{(A+C)^2-4B^2}}+1}, \\
    v_{\bm{q}} = \frac{i}{\sqrt{2}}\sqrt{\frac{A+C}{\sqrt{(A+C)^2-4B^2}}-1}.
    \label{ansatz}
\end{align}
\end{subequations}
The corresponding eigenvalues are
\begin{subequations}
\begin{align}
\begin{split}
    \omega^{\alpha}_{\bm{q}} = \frac{A-C}{2} + \frac{1}{2}\sqrt{(A+C)^2-4B^2},
\end{split}
\end{align}
and
\begin{align}
    \omega^{\beta}_{\bm{q}} = \frac{C-A}{2} + \frac{1}{2}\sqrt{(A+C)^2-4B^2}.
\end{align}
\end{subequations}
These are shown in Fig. \ref{fig:MagnonDispersionSupp}.

\section{\label{sec:ElectronMagnon}Electron-Magnon coupling}

This section considers the s-d coupling between itinerant electrons and localized spins. We consider the local coupling
\begin{align}
    H_{\mathrm{em}} = J_{\mathrm{sd}} \sum_i \bm{S}_i \cdot \bm{s}_i,
    \label{startSD}
\end{align}
where $\bm{S}_i$ is the localized spin at site $i$ and $\bm{s}_i = \sum_{\sigma, \sigma'} c_{i,\sigma}^{\dagger}\bm{\sigma}_{\sigma \sigma'} c_{i,\sigma'}$ is the spin of the itinerant electrons at site $i$.
We use the Holstein-Primakoff transformation in Eq. \eqref{HolsteinPrimakoff} to find
\begin{align}
\begin{split}
    H_{\mathrm{em}} =&\; J_{\mathrm{sd}} \sum_{i,j} \bigg[\sqrt{2S}\left(a_i c^{\dagger}_{i,\downarrow}c_{i,\uparrow} + a^\dagger_i c^{\dagger}_{i,\uparrow}c_{i,\downarrow} + b^\dagger_j c^{\dagger}_{j,\downarrow}c_{j,\uparrow} + b_j c^{\dagger}_{j,\uparrow}c_{j,\downarrow}\right) \\ &+  S \bigg((c^\dagger_{i,\uparrow}c_{i,\uparrow}+ c^\dagger_{j,\downarrow}c_{j,\downarrow})-(c^\dagger_{i,\downarrow}c_{i,\downarrow} +c^\dagger_{j,\uparrow}c_{j,\uparrow})\bigg)  \\
    &+\bigg( b_j^\dagger b_j (c^\dagger_{j,\uparrow}c_{j,\uparrow} - c^\dagger_{j,\downarrow}c_{j,\downarrow})-a_i^\dagger a_i (c^\dagger_{i,\uparrow}c_{i,\uparrow} - c^\dagger_{i,\downarrow}c_{i,\downarrow})\bigg)\bigg].
    \label{EMCoupling}
\end{split}
\end{align}
Here, the first term describes an electron spin-flip process due to a single magnon. Because this term induces a spin flip, it is restricted to scattering between altermagnetic domains in momentum space. We account for the second term in the electron Hamiltonian in Eq. \eqref{electronHamiltonian} because it will not induce any effective interaction but give rise to spin-splitting. The third term is bilinear in magnon operators. This term is often neglected. However, for altermagnets with large spin splitting, we expect the third term to be the dominant contribution to superconductivity.

We Fourier transform the electron-magnon terms in Eq. \eqref{EMCoupling} to find
\begin{widetext}
\begin{align}
\begin{split}
    H_{\mathrm{em}} =& \; J_{\mathrm{sd}} \sum_{\bm{k}, \bm{q}} \bigg[\frac{ \sqrt{2S}}{\sqrt{N}}\left(a_{\bm{q}} c^{\dagger}_{a,\bm{k}+\bm{q},\downarrow}c_{a,\bm{k},\uparrow} + a^\dagger_{\bm{q}} c^{\dagger}_{a,\bm{k}-\bm{q},\uparrow}c_{\bm{k},\downarrow} + b^\dagger_{\bm{q}} c^{\dagger}_{b,\bm{k}-\bm{q},\downarrow}c_{b,\bm{k},\uparrow} + b_{\bm{q}} c^{\dagger}_{b,\bm{k}+\bm{q},\uparrow}c_{b,\bm{k},\downarrow}\right) \\
    &+\sum_{\bm{q}'}\frac{1}{N}\bigg( b_{\bm{q}'}^\dagger b_{\bm{q}} (c^\dagger_{b,\bm{k}+\bm{q}-\bm{q}',\uparrow}c_{b,\bm{k},\uparrow} - c^\dagger_{b,\bm{k}+\bm{q}-\bm{q}',\downarrow}c_{b,\bm{k},\downarrow})-a_{\bm{q}'}^\dagger a_{\bm{q}} (c^\dagger_{a,\bm{k}+\bm{q}-\bm{q}',\uparrow}c_{a,\bm{k},\uparrow} - c^\dagger_{a,\bm{k}+\bm{q}-\bm{q}',\downarrow}c_{a,\bm{k},\downarrow})\bigg)\bigg].
\end{split}
\end{align}
In terms of band electron operators and magnon operators, the interaction is
$H_{\mathrm{em}} = H^{(1)}_{\mathrm{em}}+ H^{(2)}_{\mathrm{em}}$ with
\begin{align}
\begin{split}
    H^{(1)}_{\mathrm{em}} =& \; J_{\mathrm{sd}}\frac{ \sqrt{2S}}{\sqrt{N}} \sum_{\bm{k}, \bm{q}} \bigg[\big((\Omega^A_{\bm{k}+\bm{q},\bm{k},\downarrow, \uparrow}u_{\bm{q}}+\Omega^B_{\bm{k}+\bm{q},\bm{k},\downarrow, \uparrow}v^*_{\bm{q}}) \alpha_{\bm{q}} + (\Omega^A_{\bm{k}+\bm{q},\bm{k},\downarrow, \uparrow}v_{\bm{q}} +\Omega^B_{\bm{k}+\bm{q},\bm{k},\downarrow, \uparrow}u_{\bm{q}}^*)\beta^{\dagger}_{-\bm{q}}\big) d^\dagger_{\bm{k}+\bm{q},\downarrow}d_{\bm{k},\uparrow} \\ &+ \big((\Omega^A_{\bm{k}+\bm{q},\bm{k},\uparrow, \downarrow}u^*_{\bm{q}}+ \Omega^B_{\bm{k}+\bm{q},\bm{k},\uparrow, \downarrow}v_{\bm{q}})\alpha^\dagger_{-\bm{q}} + (\Omega^A_{\bm{k}+\bm{q},\bm{k},\uparrow ,\downarrow}v^*_{\bm{q}}+\Omega^B_{\bm{k}+\bm{q},\bm{k},\uparrow, \downarrow}u_{\bm{q}}) \beta_{\bm{q}}\big)  d^\dagger_{\bm{k}+\bm{q},\uparrow}d_{\bm{k},\downarrow}\bigg].
\end{split}
\end{align}
The second contribution is
\begin{align}
\begin{split}
    H^{(2)}_{\mathrm{em}} = &  \sum_{\bm{k}, \bm{q}, \bm{q}', \sigma}
    \frac{J_{\mathrm{sd}}\sigma}{N} \bigg( \Omega^B_{\bm{k}+\bm{q}'-\bm{q},\bm{k},\sigma,\sigma} \big(v^*_{\bm{q}}v_{\bm{q}'} \alpha_{-\bm{q}} \alpha^\dagger_{-\bm{q}'} + v^*_{\bm{q}} u_{\bm{q}'} \alpha_{-\bm{q}}\beta_{\bm{q}'} + u^*_{\bm{q}}v_{\bm{q}'}\beta_{\bm{q}}^\dagger \alpha^\dagger_{-\bm{q}'} + u^*_{\bm{q}} u_{\bm{q}'}\beta^\dagger_{\bm{q}}\beta_{\bm{q}'}\big) \\ & \; -\Omega^A_{\bm{k}+\bm{q}'-\bm{q},\bm{k},\sigma, \sigma}\big(u^*_{\bm{q}} u_{\bm{q}'} \alpha^\dagger_{\bm{q}} \alpha_{\bm{q}'} + u^*_{\bm{q}} v_{\bm{q}'} \alpha^\dagger_{\bm{q}}\beta^\dagger_{-{\bm{q}'}} + v^*_{\bm{q}}u_{\bm{q}'} \beta_{-\bm{q}}\alpha_{\bm{q}'} + v^*_{\bm{q}} v_{\bm{q}'}\beta_{-\bm{q}} \beta^\dagger_{-\bm{q}'} \big)\bigg)d^\dagger_{\bm{k}+\bm{q}'-\bm{q},\sigma}d_{\bm{k},\sigma}.
\end{split}
\end{align}
\end{widetext}
We introduced the electron coefficients $\Omega^A_{\bm{k}',\bm{k},\sigma', \sigma} \equiv q^*_{a, \bm{k}',\sigma'} q_{a, \bm{k},\sigma}$ and $\Omega^B_{\bm{k}',\bm{k},\sigma', \sigma} \equiv q^*_{b,\bm{k}',\sigma'} q_{b, \bm{k},\sigma}$. We only consider scattering on the Fermi surface, such that the band subscript $n$ is redundant.

\section{\label{sec:EffectiveInt}Effective interaction}

In this section, we derive an effective electron-electron interaction mediated by the electron-magnon coupling. We do so using a Schrieffer-Wolff transformation \cite{SchriefferWolff1966}.

Let $H_0 = H_{\mathrm{e}} + H_{\mathrm{m}}$ and $H_1 = H_{\mathrm{em}}$ such that we have $H = H_0 + \eta H_1$.
Now, we let
\begin{align}
    H'=e^{-\eta S} H e^{\eta S}
\end{align}
and expand to find
\begin{align}
\begin{split}
    H' = \; & H_0 + \eta(H_1 + \left[H_0, S\right]) + \eta^2\left[H_1, S\right] \\&+ \frac{\eta^2}{2}\left[\left[H_0, S\right], S\right] + \mathcal{O}(\eta^3).
\end{split}
\end{align}
To eliminate the linear term, we require
\begin{align}
    H_1 + \left[H_0, S\right] = 0.
    \label{condition}
\end{align}
To that end, we make the ansatz

\begin{align}
    S = S_1^A + S_1^B + \sum_{\sigma} \big(S_{2 \sigma}^A + S_{2 \sigma}^B\big),
\end{align}
with
\begin{subequations}
\begin{align}
\begin{split}
    S_1^A =& \; J_{\mathrm{sd}}\frac{ \sqrt{2S}}{\sqrt{N}} \sum_{\bm{k}, \bm{q}} 
\\ &\bigg[\Omega^A_{\bm{k}+\bm{q},\bm{k},\downarrow, \uparrow}\big(x_{\bm{k},\bm{q}}u_{\bm{q}}\alpha_{\bm{q}} + y_{\bm{k},\bm{q}}v_{\bm{q}}\beta^{\dagger}_{-\bm{q}}\big) d^\dagger_{\bm{k}+\bm{q},\downarrow}d_{\bm{k},\uparrow} \\ &+ \Omega^A_{\bm{k}+\bm{q},\bm{k},\uparrow, \downarrow}\big(z_{\bm{k},\bm{q}}u^*_{\bm{q}}\alpha^\dagger_{-\bm{q}} + w_{\bm{k},\bm{q}}v^*_{\bm{q}} \beta_{\bm{q}}\big)  d^\dagger_{\bm{k}+\bm{q},\uparrow}d_{\bm{k},\downarrow}\bigg],
\end{split}
\end{align}
\begin{align}
\begin{split}
    S_1^B =& \; J_{\mathrm{sd}}\frac{ \sqrt{2S}}{\sqrt{N}} \sum_{\bm{k}, \bm{q}} \\ &\bigg[\Omega^B_{\bm{k}+\bm{q},\bm{k},\downarrow, \uparrow}\big(x_{\bm{k},\bm{q}}v^*_{\bm{q}} \alpha_{\bm{q}} + y_{\bm{k},\bm{q}}u_{\bm{q}}^*\beta^{\dagger}_{-\bm{q}}\big) d^\dagger_{\bm{k}+\bm{q},\downarrow}d_{\bm{k},\uparrow} \\ &+ \Omega^B_{\bm{k}+\bm{q},\bm{k},\uparrow, \downarrow}\big( z_{\bm{k},\bm{q}}v_{\bm{q}}\alpha^\dagger_{-\bm{q}} + w_{\bm{k},\bm{q}}u_{\bm{q}} \beta_{\bm{q}}\big) d^\dagger_{\bm{k}+\bm{q},\uparrow}d_{\bm{k},\downarrow}\bigg],
\end{split}
\end{align}

\begin{align}
\begin{split}
    S^A_{2\sigma} =& \; \frac{J_{\mathrm{sd}}}{N}\sum_{\bm{k},\bm{q},\bm{q}'}-\sigma\big(u^*_{\bm{q}} u_{\bm{q}'} x^{A\sigma}_{\bm{k}, \bm{q}, \bm{q}'} \alpha^\dagger_{\bm{q}} \alpha_{\bm{q}'} + u^*_{\bm{q}} v_{\bm{q}'} y^{A\sigma}_{\bm{k}, \bm{q}, \bm{q}'}\alpha^\dagger_{\bm{q}}\beta^\dagger_{-\bm{q}'} \\&+ v^*_{\bm{q} }u_{\bm{q}'} z^{A\sigma}_{\bm{k}, \bm{q}, \bm{q}'}\beta_{-\bm{q}}\alpha_{\bm{q}'} + v^*_{\bm{q}} v_{\bm{q}'}w^{A\sigma}_{\bm{k}, \bm{q}, \bm{q}'}\beta_{-\bm{q}} \beta^\dagger_{-\bm{q}'}\big) \\
    &\times (\Omega^A_{\bm{k} + \bm{q}-\bm{q}',\bm{k},\sigma, \sigma}d^\dagger_{\bm{k} + \bm{q} -\bm{q}',\sigma}d_{\bm{k},\sigma}),
\end{split}
\end{align}
and
\begin{align}
\begin{split}
    S^B_{2\sigma} =& \;  \frac{J_{\mathrm{sd}}}{N}\sum_{\bm{k},\bm{q},\bm{q}'}\sigma\big(v^*_{\bm{q}}v_{\bm{q}'} x^{B\sigma}_{\bm{k},\bm{q}, \bm{q}'}\alpha_{-\bm{q}} \alpha^\dagger_{-\bm{q}'} + v^*_{\bm{q}} u_{\bm{q}'} z^{B\sigma}_{\bm{k}, \bm{q},\bm{q}'}\alpha_{-\bm{q}}\beta_{\bm{q}'} \\ &+ u^*_{\bm{q}}v_{\bm{q}'}y^{B\sigma}_{\bm{k}, \bm{q},\bm{q}'}\beta_{\bm{q}}^\dagger \alpha^\dagger_{-\bm{q}'} + u^*_{\bm{q}} u_{\bm{q}'}w^{B\sigma}_{\bm{k}, \bm{q},\bm{q}'}\beta^\dagger_{\bm{q}}\beta_{\bm{q}'}\big) \\ &\times ( \Omega^B_{\bm{k}+\bm{q}-\bm{q}',\bm{k},\sigma, \sigma} d^\dagger_{\bm{k}+\bm{q}-\bm{q}',\sigma}d_{\bm{k},\sigma}).
\end{split}
\end{align}
\end{subequations}

The condition in Eq. \eqref{condition} determines the parameters
\begin{subequations}
\begin{align}
     x_{\bm{k}, \bm{q}} =   \frac{1}{\varepsilon_{\bm{k},\uparrow} -\varepsilon_{\bm{k}+\bm{q},\downarrow} + \omega^{\alpha}_{\bm{q}}}, \\
    y_{\bm{k}, \bm{q}} = \frac{1}{\varepsilon_{\bm{k},\uparrow} -\varepsilon_{\bm{k}+\bm{q},\downarrow} - \omega^{\beta}_{\bm{q}}}, \\
    z_{\bm{k}, \bm{q}} =   \frac{1}{\varepsilon_{\bm{k},\downarrow} -\varepsilon_{\bm{k}+\bm{q},\uparrow} - \omega^{\alpha}_{\bm{q}}}, \\
    w_{\bm{k}, \bm{q}} = \frac{1}{\varepsilon_{\bm{k},\downarrow} -\varepsilon_{\bm{k}+\bm{q},\uparrow} + \omega^{\beta}_{\bm{q}}},
\end{align}
\end{subequations}
and
\begin{subequations}
\begin{align}
    x^{A\sigma}_{\bm{k}, \bm{q}, \bm{q}'} = \;& \frac{1}{\varepsilon_{\bm{k},\sigma} - \varepsilon_{\bm{k}+\bm{q} - \bm{q}', \sigma} - \omega^\alpha_{\bm{q}} + \omega^\alpha_{\bm{q}'}}, \\
    y^{A\sigma}_{\bm{k}, \bm{q}, \bm{q}'} = \;& \frac{1}{\varepsilon_{\bm{k},\sigma} - \varepsilon_{\bm{k}+ \bm{q} - \bm{q}', \sigma} - \omega^\alpha_{\bm{q}} - \omega^\beta_{-\bm{q}'}},
    \\
    z^{A\sigma}_{\bm{k}, \bm{q}, \bm{q}'} = \;& \frac{1}{\varepsilon_{\bm{k},\sigma} - \varepsilon_{\bm{k}+ \bm{q} - \bm{q}', \sigma} + \omega^\alpha_{\bm{q}'} + \omega^\beta_{-\bm{q}}},
    \\
    w^{A\sigma}_{\bm{k}, \bm{q}, \bm{q}'} = \;& \frac{1}{\varepsilon_{\bm{k},\sigma} - \varepsilon_{\bm{k}+ \bm{q} - \bm{q}', \sigma} - \omega^\beta_{-\bm{q}'} + \omega^\beta_{-\bm{q}}},
\end{align}
\end{subequations}
\begin{subequations}
\begin{align}
    x^{B\sigma}_{\bm{k}, \bm{q}, \bm{q}'} = \;& \frac{1}{\varepsilon_{\bm{k},\sigma} - \varepsilon_{\bm{k}+\bm{q} - \bm{q}', \sigma} - \omega^\alpha_{-\bm{q}'} + \omega^\alpha_{-\bm{q}}}, \\
    y^{B\sigma}_{\bm{k}, \bm{q}, \bm{q}'} = \;& \frac{1}{\varepsilon_{\bm{k},\sigma} - \varepsilon_{\bm{k}+ \bm{q} - \bm{q}', \sigma} - \omega^\alpha_{-\bm{q}'} - \omega^\beta_{\bm{q}}},
    \\
    z^{B\sigma}_{\bm{k}, \bm{q}, \bm{q}'} = \;& \frac{1}{\varepsilon_{\bm{k},\sigma} - \varepsilon_{\bm{k}+\bm{q} - \bm{q}', \sigma} + \omega^\alpha_{-\bm{q}} + \omega^\beta_{\bm{q}'}},
    \\
    w^{B\sigma}_{\bm{k}, \bm{q}, \bm{q}'} = \;& \frac{1}{\varepsilon_{\bm{k},\sigma} - \varepsilon_{\bm{k}+\bm{q} - \bm{q}', \sigma} - \omega^\beta_{\bm{q}} + \omega^\beta_{\bm{q}'}}.
\end{align}
\end{subequations}
The $C_{2z}$ symmetry in Eq. \eqref{eq:sym1} leads to 
$\omega_{-\bm{q}}=\omega_{\bm{q}}$. The case $\omega^{\alpha}_{\bm{q}} \neq \omega^{\beta}_{\bm{q}}$ distinguishes these parameters from their analogs in conventional antiferromagnets.

The effective interaction is
\begin{align}
    H_{\mathrm{eff}} = \frac{\eta^2}{2}\left[H_1, S\right].
\end{align}
This commutator gives rise to several terms. We consider the quartic terms in electron operators because these terms constitute an effective electron-electron interaction. We consider scattering processes on the Fermi surface between electrons of opposite momentum only. The resulting interaction is
\begin{align}
\begin{split}
    H_{\mathrm{e-e}} =& \sum_{\bm{k}, \bm{k}'} V^{(1)}_{\bm{k}, \bm{k}' }d^\dagger_{\bm{k}',\uparrow}d^\dagger_{-\bm{k}',\downarrow}d_{-\bm{k},\downarrow}d_{\bm{k},\uparrow}   \\& + \sum_{\bm{k}, \bm{k}', \sigma} V^{(2)}_{\bm{k}, \bm{k}', \sigma} d^\dagger_{\bm{k}',\sigma}d^\dagger_{-\bm{k}',\sigma}d_{-\bm{k},\sigma} d_{\bm{k},\sigma},
\end{split}
\end{align}
with
\begin{widetext}
\begin{align}
\begin{split}
    V^{(1)}_{\bm{k}, \bm{k}'} =\;&  \frac{J^2_{\mathrm{sd}}S}{N}\bigg[\frac{1}{\omega^\alpha_{\bm{k}+\bm{k}'}}\bigg(\Omega^A_{\bm{k}',\bm{k},\uparrow, \downarrow}\Omega^A_{\bm{k}',\bm{k},\downarrow, \uparrow}\abs{u_{\bm{k}'+\bm{k}}}^2 + \Omega^B_{\bm{k}',\bm{k},\uparrow, \downarrow}\Omega^B_{\bm{k}',\bm{k},\downarrow, \uparrow}\abs{v_{\bm{k}'+\bm{k}}}^2\bigg) \\
    & \; + \frac{1}{\omega^\beta_{\bm{k}+\bm{k}'}}\bigg(\Omega^A_{\bm{k}',\bm{k},\uparrow, \downarrow}\Omega^A_{\bm{k}',\bm{k},\downarrow, \uparrow}\abs{v_{\bm{k}'+\bm{k}}}^2 + \Omega^B_{\bm{k}',\bm{k},\uparrow, \downarrow}\Omega^B_{\bm{k}',\bm{k},\downarrow, \uparrow}\abs{u_{\bm{k}'+\bm{k}}}^2\bigg) \\
    & \;+ \bigg(\frac{1}{\omega^\alpha_{\bm{k}+\bm{k}'}} + \frac{1}{\omega^\beta_{\bm{k}+\bm{k}'}}\bigg) u_{\bm{k}+\bm{k}'}v_{\bm{k}+\bm{k}'}\big(\Omega^A_{\bm{k}',\bm{k},\uparrow, \downarrow}\Omega^B_{\bm{k}', \bm{k},\downarrow, \uparrow} + \Omega^B_{\bm{k}', \bm{k},\uparrow, \downarrow}\Omega^A_{\bm{k}',\bm{k},\downarrow, \uparrow}\big)\bigg],
    \label{V1}
\end{split}
\end{align}
and
\begin{align}
\begin{split}
    V^{(2)}_{\bm{k}, \bm{k}', \sigma} = \frac{J^2_{\mathrm{sd}}}{N^2}\sum_{\bm{Q}} \bigg(\Omega^A_{\bm{k}',\bm{k},\sigma, \sigma}
     \abs{v_{\frac{\bm{Q}-(\bm{k}'-\bm{k})}{2}}} \abs{u_{\frac{\bm{Q}+(\bm{k}'-\bm{k})}{2}}} - \Omega^B_{\bm{k}',\bm{k},\sigma, \sigma}
     \abs{v_{\frac{\bm{Q}+(\bm{k}'-\bm{k})}{2}}} \abs{u_{\frac{\bm{Q}-(\bm{k}'-\bm{k})}{2}}}\bigg)^2 y^{A\sigma}_{\bm{k}, \frac{\bm{Q}+(\bm{k}'-\bm{k})}{2},\frac{\bm{Q}-(\bm{k}'-\bm{k})}{2}}.
    \label{V2A}
\end{split}
\end{align}
\end{widetext}
Here, the first term is an interaction between electrons of opposite spins. This term can not give rise to zero momentum Cooper pairs for spin-polarized bands and, thus, not a robust superconducting state. However, the interaction in Eq. \eqref{V1} could be relevant for weak altermagnets. In this case, there is no $\mathcal{PT}$-symmetry relating $\Omega^A$ and $\Omega^B$ as for conventional antiferromagnets. This is interesting because it could lead to an intrinsic squeezing-enhanced superconducting state. For conventional antiferromagnets, this behavior is reserved for uncompensated interfaces \cite{Erlandsen2019PRB, Erlandsen2021PRB}. In the second term, we introduced $\bm{Q} = \bm{q} + \bm{q}'$. In the following, we consider potential superconductivity arising from the interaction in Eq. \eqref{V2A}.

\section{\label{sec:Superconductivity}Superconductivity}

To estimate the critical temperature and the $\bm{k}$-dependence of superconducting gap $\Delta_{\bm{k}, \sigma}$, we consider the linearized gap equation
\begin{align}
\begin{split}
    \Delta_{\bm{k},\sigma}
    = -\sum_{\bm{k}'}V_{\bm{k},\bm{k}',\sigma}\frac{\Delta_{\bm{k}',\sigma}}{\varepsilon_{\bm{k}',\sigma}}\tanh{\left(\frac{\beta \varepsilon_{\bm{k}',\sigma}}{2}\right)}.
\end{split}
\end{align}
In the continuum limit, the sum is equivalent to an integral over the Brillouin zone
\begin{align}
\begin{split}
    \Delta_{\bm{k},\sigma}
    = -\frac{N}{A_{\mathrm{BZ}}}\int_{\mathrm{BZ}}d\bm{k}'V_{\bm{k},\bm{k}',\sigma}\frac{\Delta_{\bm{k}',\sigma}}{\varepsilon_{\bm{k}',\sigma}}\tanh{\left(\frac{\beta \varepsilon_{\bm{k}',\sigma}}{2}\right)}.
\end{split}
\end{align}
We now assume that the effective interaction and the gap are constant perpendicular to the Fermi surface for $\abs{\varepsilon_{\bm{k}}} < \omega_M$ and zero otherwise. The integral separates into one part perpendicular to the Fermi surface, and one parallel to the Fermi surface. We get
\begin{align}
\begin{split}
    \Delta_{\bm{k}_\parallel,\sigma} =& -\frac{2N}{A_{\mathrm{BZ}}} \int d \bm{k}'_{\parallel}\abs{\frac{d\varepsilon}{d \bm{k}'_\perp}}^{-1} V_{\bm{k}_\parallel, \bm{k}'_\parallel, \sigma}\Delta_{\bm{k}'_\parallel, \sigma}\\ &\times \int_0^{\omega_M} d\varepsilon \frac{\tanh{\beta\varepsilon/2}}{\varepsilon},
\end{split}
\end{align}
Here, $\omega_M$ is the magnon cut-off frequency. We solve for the largest eigenvalue of the equation
\begin{align}
     -\frac{2N}{A_{\mathrm{BZ}}} \int d \bm{k}'_{\parallel}\abs{\frac{d\varepsilon}{d \bm{k}'_\perp}}^{-1} V_{\bm{k}_\parallel, \bm{k}'_\parallel,\sigma}\Delta_{\bm{k}'_\parallel, \sigma} = \lambda \Delta_{\bm{k}_\parallel, \sigma}.
\end{align}
Equivalently, we have
\begin{align}
    -\frac{2N S_{\mathrm{FS}}}{N_{\mathrm{samp}}A_{\mathrm{BZ}}} \sum_{\bm{k}'_\parallel} \abs{\frac{d\varepsilon}{d \bm{k}'_\perp}}^{-1} V_{\bm{k}_\parallel, \bm{k}'_\parallel, \sigma}\Delta_{\bm{k}'_\parallel, \sigma} = \lambda \Delta_{\bm{k}_\parallel, \sigma}.
    \label{matrixEquation}
\end{align}
The number of points $N$ in the Brillouin zone in the numerator cancels to the denominator in $V_{\bm{k}, \bm{k}', \sigma}$. The area of the Brillouin zone is $A_{\mathrm{BZ}}=\pi^2$. We sample the Fermi surface uniformly at intervals given by Eq. \eqref{sampleInterval} with a total of $N_{\mathrm{samp}}$ points.
Now, Eq. \eqref{matrixEquation} has multiple eigenvalues. We pick out the largest one because it determines the critical temperature. 
The corresponding eigenvector gives the $\bm{k}$-dependence of the gap $\Delta_{\bm{k},\sigma}$.
We get
\begin{align}
    1 = \lambda_{\mathrm{eff}} \int_0^{\omega_M} d\varepsilon \frac{\tanh{\beta_c\varepsilon/2}}{\varepsilon},
\end{align}
where $\beta_c$ is the inverse critical temperature.
The critical temperature is then related to the largest eigenvalue shown in Eq. \eqref{CriticalTemperature}.

\section{\label{sec:orbitalOrdering}Altermagnetism due to ordering of local orbitals}
The main text considers the interplay between magnetic and non-magnetic sites as an origin of altermagnetic properties.
Anisotropic ordering of local orbitals is a different mechanism that can potentially induce an altermagnetic state.

\begin{figure}[t!]
\centering
\begin{subfloat}[\label{subfig:orbitalLattice}]{\includegraphics[width=0.38\columnwidth]{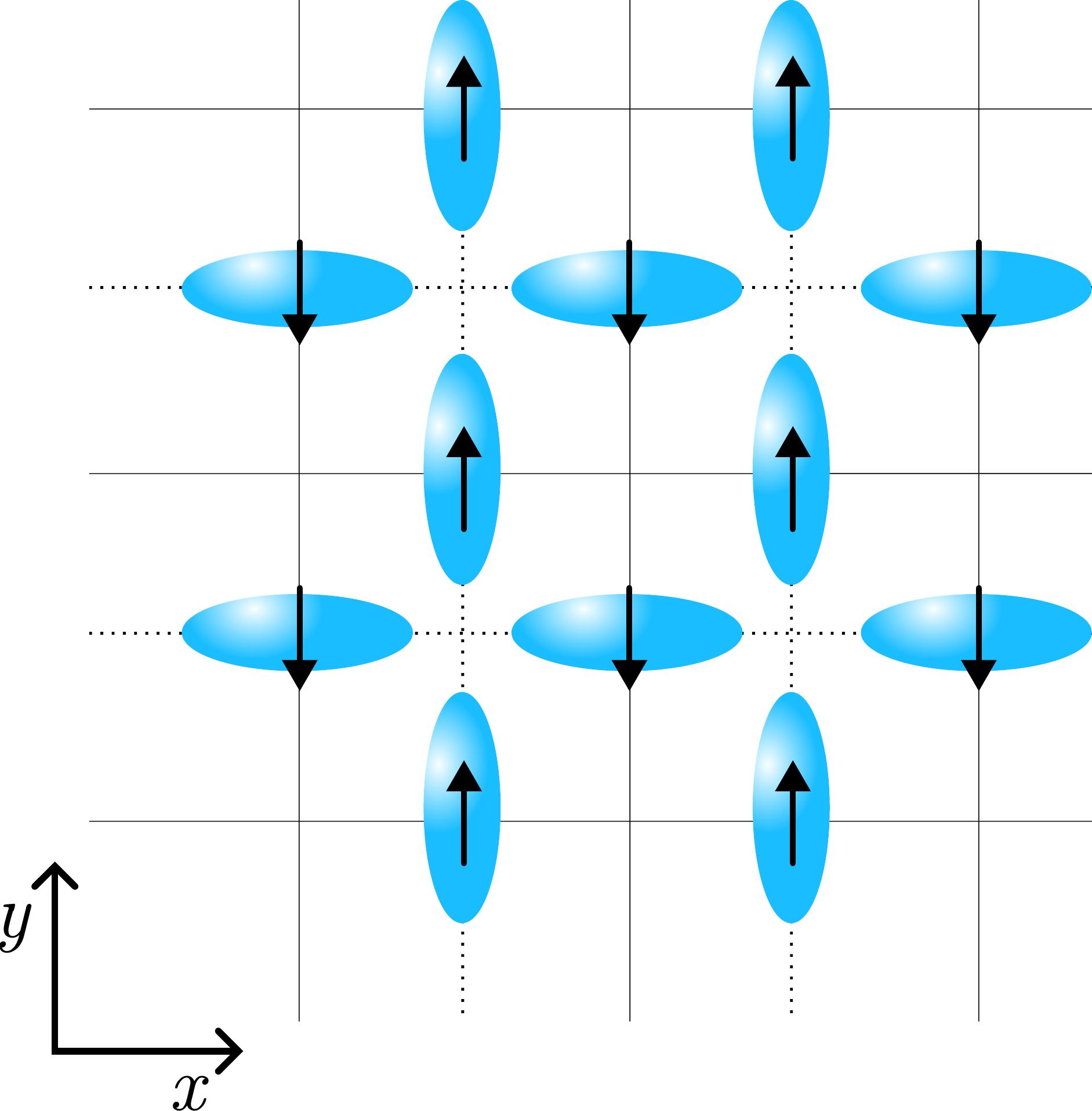}}
\end{subfloat}
\begin{subfloat}[\label{subfig:orbitalElectrons}]{\includegraphics[width=0.60\columnwidth]{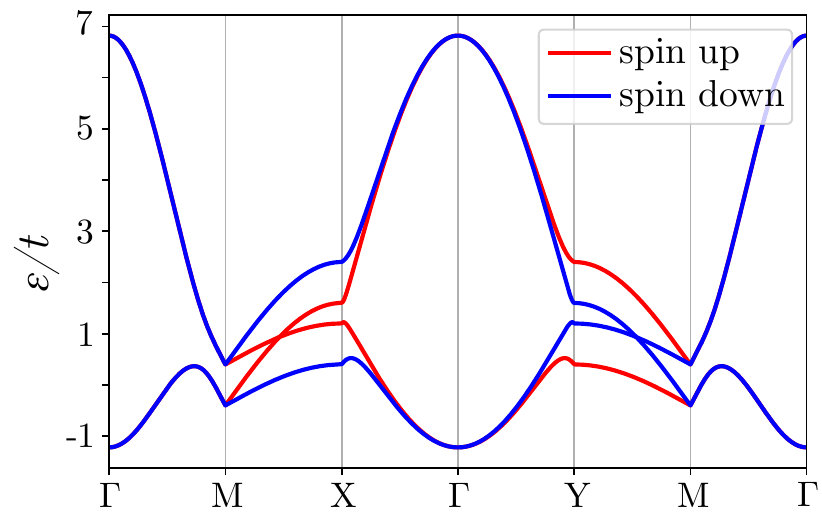}}
\end{subfloat}
\caption{\label{fig:OrbitalElectronBands}a) A two-dimensional altermagnet, with lattice sites that reflect an ordering of local atomic- or molecular orbitals. The orbital ordering breaks the $P\mathcal{T}$-symmetry to allow for altermagnetic properties. b) The electron dispersion for the tight-binding model in Eq. \eqref{electronOrbitalHamiltonian} with $t'_2/t = 1$, $t_2/t = 0.4$, $J_{\mathrm{sd}}S/t = 0.4$ and $\mu/t=0$. The bands are spin-split due to the anisotropic hopping parameter.}
\end{figure}

In this section, we briefly consider the mechanism of orbital ordering as the origin of altermagnetism. To that end, we employ a tight-binding model with anisotropic hopping parameters on the lattice in Fig. \ref{subfig:orbitalLattice}.

We capture the electron properties of the orbital altermagnet through the tight-binding model
\begin{align}
\begin{split}
    H_e =& \; t_d\sum_{\left<i,j\right>,\sigma}  c_{i,\sigma}^{\dagger} c_{j,\sigma}  + t_2\sum_{\left<\left<i,j\right>\right>, \sigma}  c_{i,\sigma}^{\dagger} c_{j,\sigma} +  t'_2\sum_{\left<\left<i,j\right>\right>, \sigma}  c_{i,\sigma}^{\dagger} c_{j,\sigma}
  \\ &- J_{\mathrm{sd}}\sum_{i, \sigma, \sigma'} \bm{S}_i\cdot c_{i,\sigma}^{\dagger}\bm{\sigma}_{\sigma \sigma'} c_{i,\sigma'} -\mu \sum_{i, \sigma}  c_{i,\sigma}^{\dagger} c_{i,\sigma}.
    \label{electronOrbitalHamiltonian}
\end{split}
\end{align}
Here, $t_d$ is the hopping parameter for diagonal hopping, $t_2$ and $t'_2$ denote hopping along the solid and dotted lines in Fig. \ref{subfig:orbitalLattice}), respectively.

The electron eigenvalues are
\begin{subequations}
\begin{align}
    \varepsilon_1(\bm{k}) = A_e+C_e -\sqrt{4B_e^2+(C_e-A_e \pm J_{\mathrm{sd}}S)^2}-\mu, \\
    \varepsilon_2(\bm{k}) = A_e+C_e + \sqrt{4B_e^2+(C_e-A_e \pm J_{\mathrm{sd}}S)^2}- \mu,
\end{align}
\label{eigenvaluesOrbital}
\end{subequations}
where, $(+)$ is for the spin-up bands and $(-)$ is for the spin-down bands. The parameters are defined as
\begin{subequations}
\begin{align}
    A_e(\bm{k}) =\;& t_2\cos{(k_x a)} +  t'_2\cos{(k_y a)}, \\
    B_e(\bm{k}) =\;& t\big(\cos{(k_x a + k_y a)} + \cos{(k_x a - k_y a)}\big), \\
    C_e(\bm{k}) =\;& t_2\cos{(k_y a)} +  t'_2\cos{(k_x a)}.
\end{align}
\end{subequations}
The spin-splitting can be found directly from Eq. \eqref{eigenvaluesOrbital} for both bands. It is
\begin{align}
\begin{split}
    \Delta_{ss}(\bm{k}) = \pm \bigg(\sqrt{4B_e^2+(C_e-A_e + J_{\mathrm{sd}}S)^2} \\- \sqrt{4B_e^2+(C_e-A_e - J_{\mathrm{sd}}S)^2}\bigg),
\end{split}
\end{align}
where, we use $(+)$ for the upper bands and $(-)$ for the lower bands.




\bibliography{apssamp}

\end{document}